\documentclass[seceq]{ptptex}
\usepackage{graphicx}
\usepackage{wrapft}

\usepackage{amsmath}
\usepackage{bm}
\usepackage{url}

\newcommand{\lsim}{\mbox{\raisebox{-0.6ex}{$\stackrel{<}{\sim}$}}\:}
\newcommand{\RE}{\mbox{${\rm Re}$}}
\newcommand{\Tr}{\mbox{${\rm Tr}$}}

\title{
Dynamical modeling of high energy heavy ion collisions%
}

\author{
Tetsufumi \textsc{Hirano}$^{1,2}$ and %
Yasushi \textsc{Nara}$^{3}$%
}

\inst{
$^{1}$Department of Physics, Sophia University, Tokyo 102-8554, Japan\\
$^{2}$Department of Physics, the University of Tokyo, Tokyo 113-0033, Japan\\
$^{3}$Akita International University, Yuwa, Akita-city 010-1292, Japan
}

\abst{
We present theoretical approaches to high energy nuclear collisions
in detail putting a special emphasis on
technical aspects of numerical simulations.
Models include relativistic hydrodynamics, Monte-Carlo implementation
of $k_T$-factorization formula, jet quenching in expanding fluids,
a hadronic transport model
and the Vlasov equation for colored particles.
}

\begin{document}

\maketitle

\section{Introduction}
\label{s:intro}

One of the primary purposes of heavy ion physics at ultrarelativistic energies
is to explore 
properties of strongly interacting matter at high temperature and densities,
namely, the quark gluon plasma (QGP)~\cite{Gyulassy:2004vg,sQGP,Yagi:2005yb}.
To this purpose, collider experiments of high energy $pp$, $dA$ and $AA$ collisions
have been performed at both Relativistic Heavy Ion Collider (RHIC)
in Brookhaven National Laboratory (BNL) and Large Hadron Collider (LHC)
in European Organization for Nuclear Research (CERN).
It is of particular importance to comprehensively understand space-time evolution
of matter created in high energy heavy ion collisions so that
one can extract information
on the detailed properties of matter from experimental data.

High energy nuclear collisions contain rich physics
and exhibit many aspects of dynamics
according to relevant energy and
time scales.
Collisions of two energetic nuclei can be viewed as those of
highly coherent dense gluons.
Universal behaviors of hadrons and nuclei at the high energy limit
are called the color glass condensate (CGC).~\cite{cgc}
Just after the collision, longitudinal color electric and magnetic
fields between the two passing nuclei, which is called color flux tubes,\cite{Gatoff:1987uf}
are produced from CGC initial conditions.
Subsequent evolution of these flux tubes is called ``glasma"~\cite{Lappi:2006fp}.
Long range rapidity correlations
in the glasma~\cite{Dumitru:2008wn,Dusling:2009ni,Dumitru:2010iy}
provide a natural explanation for the ridge structure
observed at RHIC~\cite{:2009qa,Alver:2009id} and
LHC~\cite{Khachatryan:2010gv}.
It was pointed out that
instabilities of the gluon fields could
play a significant role in the process of thermalization.~\cite{Mrowczynski,RomStri,ALM}

A model based on 
$k_T$-factorization formula~\cite{KLN,fKLN,Gelis:2006tb,Albacete:2007sm,Levin:2011hr,
Tribedy:2010ab,MCKLN,mckt,Dumitru:2011wq}
or classical Yang-Mills approach~\cite{KV,Lappi} reproduces
the multiplicity distribution for charged hadrons
at RHIC and LHC.
However, initial transverse energy per particle is large compared
with the experimental data~\cite{KNV3,Lappi,mckt}.
Glasma evolution based on a 2+1 dimensional classical Yang-Mills
simulation does not account for
elliptic flow data at RHIC~\cite{KNV2}. These facts suggest
the necessity of inclusion of further evolution with much stronger
interaction, \textit{e.g.}, hydrodynamical evolution.
Indeed, a nearly perfect fluid
picture~\cite{Kolb:2000fha,Huovinen:2001cy,Teaney:2000cw,Hirano:2001eu,Hirano:2002ds}
turns out to explain large
elliptic flow \cite{Ollitrault} observed at 
RHIC \cite{STARv2,PHENIXv2,PHOBOSv2} and
LHC,\cite{Aamodt:2010pa,ATLAS:2011ah,Velkovska:2011zz}
which leads to establishment of a new paradigm ``strongly coupled QGP (sQGP)"
\cite{Gyulassy:2004vg,sQGP,BNL,Hirano:2005wx}.
These hydrodynamic simulations require
very short  ($\lsim 1$ fm/$c$) thermalization time.

Due to expansion and cooling down, a QGP fluid
becomes eventually a hadronic gas at a late stage whose evolution can be described 
by hadronic transport
 models~\cite{Bass:1999tu,Teaney:2000cw,Hirano:2005xf,Nonaka:2006yn,Werner:2010aa}.
Realistic gradual freeze-out both chemically and kinetically
can be naturally treated in these models.
It is claimed that a QGP fluid picture and a hadron gas picture
are demanded to understand $p_{T}$ speactra and differential elliptic flow parameter
for identified hadrons simultaneously.\cite{Hirano:2005wx}
For the hadronization processes, it has been claimed
that quark recombination/coalescence is
important for intermediate transverse momentum region.\cite{Fries:2008hs}

On the other hand, high momentum jets are also created in the 
collision at collider energies.
These jets have to traverse the matter
in heavy ion collisions and, therefore, interaction
with the expanding matter should be treated accordingly.
During traveling through the medium, jets interact with soft
matter and lose their energies (jet quenching).
Therefore, high transverse momentum hadrons are good probes of
the bulk matter~\cite{Wiedemann:2009sh,d'Enterria:2009am,Majumder:2010qh,Fries:2010ht}.
Suppression of high transverse momentum hadrons
was observed at RHIC~\cite{jetqPHENIX,jetqSTAR} and LHC~\cite{jetqLHC}.
Disappearance of the away-side peak in azimuthal correlation functions
for high transverse momentum hadrons is also observed in central Au+Au
collisions at RHIC.\cite{Adler:2002tq} 
High $p_{T}$ suppression in heavy ion collisions
turns out to be attributed to the final state interaction
because of the Cronin enhancement
and the existence of back-to-back
correlation in d+Au collisions at midrapidity at RHIC.\cite{RHICdAu}
Theoretical approaches to describe such jet quenching need space-time
evolution of parton density through a trajectory of a jet.
Hydrodynamical simulation provides
such a parton density. 
A hydro + jet model was proposed in Refs.~\citen{Hirano:2002sc,Hirano:2003hq}
later followed by Refs.~\citen{Renk:2006sx,Bass:2008rv,Schenke:2009gb}
incorporating more realistic models for parton energy loss in a dense medium.

Recent experimental data on higher order anisotropic
flow~\cite{Adare:2011tg,ALICE:2011ab,Aad:2012bu}
call attention to
the importance of initial state fluctuations in heavy ion collisions.
It is known that event-by-event fluctuations~\cite{Alver:2010gr}
lead to the higher order anisotropic flow such as triangular flow
quantified by third harmonic component of azimuthal angle distributions,
and importance of event-by-event simulations with fluctuating
initial conditions  has been 
realized~\cite{Werner:2010aa,Andrade:2006yh,Alver:2010dn,Petersen:2010cw,Qin:2010pf,Schenke:2010rr,Qiu:2011iv}.
The triangular flow contributes also most of
the ridge observed in the two-particle azimuthal correlation.
So far only fluctuations from configuration of nucleons inside a colliding nucleus
are included in the hydrodynamical simulations.
Recently fluctuations from particle production itself
are studied~\cite{Tribedy:2010ab} by including negative binomial distribution
which was obtained by the glasma~\cite{Gelis:2009wh}.
It was found that multiplicity fluctuations in $pp$ and $pA$ collisions
at midrapidity exhibit Koba-Nielsen-Olsen (KNO) scaling~\cite{Dumitru:2012yr}.
Fluctuations of particle production increase higher order anisotropy
such as triangularity by about 50\%\cite{Dumitru:2012yr}.
DIPSY event generator which includes the fluctuations arising
from dipole evolution also predicts
larger higher order anisotropy~\cite{Flensburg:2011wx}.
Viscous hydrodynamic simulations with fluctuating Glasma initial conditions
have been performed recently~\cite{Schenke:2012wb}.
Monte-Carlo Glauber type model in which effects of nucleon-nucleon
correlations~\cite{Alvioli:2009ab,Alvioli:2011sk}
are included was proposed to construct more realistic 
nuclear configurations for initial conditions of subsequent evolution.

\begin{figure}
\begin{center}
\includegraphics[width=3.4in]{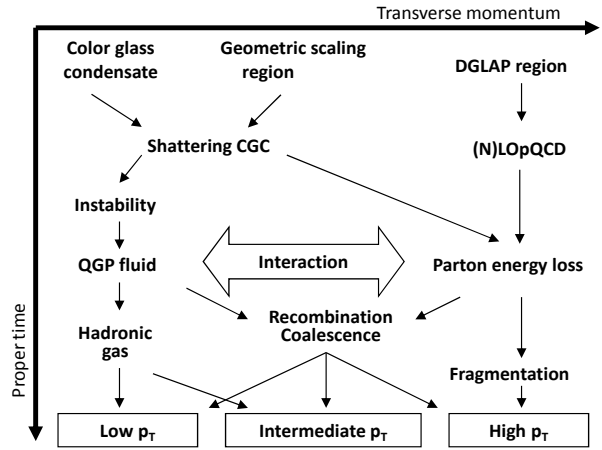}
\caption{Dynamical modeling of relativistic heavy ion collisions
in view from proper time and energy scale.}
\label{fig:intro}
\end{center}
\end{figure}

As described above, it is needed to
incorporate all such different physics consistently
to have unified and better understanding of the space-time evolution
of the system created in high energy nuclear collisions.
Figure \ref{fig:intro} shows several important aspects of dynamics of
relativistic heavy ion collisions 
according to time and energy scales.
Experimental observables reflect all the history of evolution of matter
staring from initial colliding nuclei to final free-streaming hadrons.
A first attempt to an integrated approach to the heavy ion collision as a whole
was done in Ref.~\citen{Hirano:2004en} in which full three dimensional
ideal hydrodynamic
simulations with initial conditions taken from a CGC based model were performed and
parton energy loss was simulated in these expanding fluids.
In this review, we present technical and numerical aspects of
these important modules for high energy
heavy ion collisions; CGC, relativistic hydrodynamics, parton energy loss,
hadronic transport models and Vlasov model for colored particles.

This paper is organized as follows. In Sec.~\ref{s:hydroec2},
we explain in detail numerical aspects of relativistic ideal hydrodynamics
in heavy ion collisions.
In Sec.~\ref{sec:initialcondition},
Monte-Carlo implementation of the CGC initial conditions
based on the $k_T$-factorization formula is discussed.
Energy loss of energetic partons
in an expanding QGP fluid is briefly discussed in
Sec.~\ref{sec:energyloss}.
In Sec.~\ref{sec:transport},
a cascade method and cross sections in the hadronic transport model JAM
are briefly summarized.
In Sec.~\ref{sec:nonabelian},
we discuss how to solve the Vlasov equation for colored particles
by employing the particle-in-cell method.
The final section is devoted to summary and conclusion.

\section{Relativistic Hydrodynamics}
\label{s:hydroec2}

Relativistic hydrodynamics is one of the key dynamical framework
to describe the space-time evolution of matter created in relativistic
heavy ion collisions.
Since the main goal in the physics of
relativistic heavy ion collisions is to understand the properties
of matter under (local) equilibrium,
one can apply hydrodynamics,
in which local thermal equilibrium is assumed,
to dynamical description
of created matter in any cases as a bottom-up approach
to see whether the hydrodynamic description works well.
In this section, we briefly overview framework 
 of relativistic ideal hydrodynamics and its
numerical aspects.

\subsection{Relativistic ideal hydrodynamics}
Relativistic hydrodynamic equations describe conservation laws
of energy and momentum
\begin{eqnarray}
\label{eq:hydroeq}
\partial_\mu T^{\mu \nu}(x) & = & 0,
\end{eqnarray}
together with the conservation of charges
\begin{eqnarray}
\label{eq:chargecons}
\partial_\mu N^{\mu}_{i}(x) & = & 0.
\end{eqnarray}
Here $T^{\mu \nu}$ is the energy momentum tensor
and $N^{\mu}_{i}$ is the $i$-th conserved current.
With an assumption of ideal hydrodynamics
where all dissipative effects are neglected,
 one can decompose the energy momentum tensor and the conserved currents as
\begin{eqnarray}
T^{\mu \nu} & = & e u^\mu u^\nu - P (g^{\mu \nu} - u^\mu u^\nu),\\
\label{eq:tmunu}
N^{\mu}_{i} & = & n_{i} u^{\mu}, 
\end{eqnarray}
where $e$, $P$, $n_{i}$ and $u^\mu = \gamma (1, \bm{v})= \frac{1}{\sqrt{1-v^2}}(1, \bm{v})$ are
energy density, pressure, $i$-th conserved charge density and four flow velocity, respectively.
Minkowski metric in this paper is defined as $g^{\mu \nu} = $ diag$(1, -1, -1, -1)$.
Instead of Eqs.~(\ref{eq:hydroeq}) and (\ref{eq:chargecons}),
the following expression of the balance equations might be also
convenient when compared with non-relativistic equations:
\begin{eqnarray}
&& \frac{\partial}{\partial t}E + \nabla \cdot (E+P)\bm{v} = 0,\\
&& \frac{\partial}{\partial t}M^{k} + \nabla \cdot M^{k} \bm{v} = -\nabla^{k}P,\\
&& \frac{\partial}{\partial t}N_{i} + \nabla \cdot N_{i} \bm{v} = 0,
\label{eq:baryoncons}
\end{eqnarray}
where 
\begin{eqnarray}
E & = & (e+P) \gamma^2 -P,\\
\bm{M} & = & (e+P) \gamma^2 \bm{v},\\
N_{i} & = & n_{i} \gamma.
\end{eqnarray}

In ideal hydrodynamic framework,
the equation of state plays an important role.
First, the hydrodynamic equations (\ref{eq:hydroeq}) and (\ref{eq:chargecons})
are not closed as a system of partial differential equations:
the number of unknowns is 6 (energy density, pressure, charge density
and three components of flow velocity) in the case of one conserved charge
while the number of equations is 5.
So the system is closed when a relation among unknowns, \textit{e.g.}, $P=P(e, n)$
is specified.
Second, equations~(\ref{eq:hydroeq}) and (\ref{eq:chargecons}) 
just describe conservation laws and are the so-called balance equations.
Under an assumption of
local thermal equilibrium,
one can utilize the equation of state $P=P(e, n)$ which only reflects
the microscopic dynamics.
Since the collective flow is generated by pressure gradient,
it is sensitive to the equation of state and the degree of kinetic equilibrium.
This is one of the reasons why the collective flow has been focused in relativistic
heavy ion collisions.

The equation of state can be taken from results from the first principle
calculations of QCD thermodynamics, namely, lattice QCD.
\cite{Cheng:2007jq,Bazavov:2009zn,Borsanyi:2010cj,Umeda:2012er}
A few comments regarding this are in order here.

\begin{enumerate}
\item In the current status of the lattice QCD results,
the equation of state is not so reliable in the low temperature region ($T\lsim 100$ MeV).
Then one can connect lattice QCD results at high temperature
with results from
the hadronic resonance gas model at low temperature.
\cite{Huovinen:2009yb,EoSsite}
Practically, this is also demanded from a point of view of freezeout
at which all hydrodynamic variables are switched to a particle picture
employing the Cooper-Frye formula.\cite{Cooper:1974mv}
Energy, momentum and charges
are conserved in this formula only when the hadronic resonance gas picture is valid.

\item  Monte-Carlo calculations of lattice QCD
at finite baryonic chemical
potential \cite{Levkova:2012jd}
suffer from a severe issue due to the so-called sign problem.
The equation of state at finite baryon density, however,
would be demanded in lower collision energies or in forward/backward
rapidity regions.

\item Even if results will become reliable in the low temperature region 
and/or finite baryon density region in lattice QCD,
there is an issue on chemical freezeout since
all thermodynamic variables are obtained for thermally as well as chemically equilibrated
states. In the actual hadronic matter created in relativistic
heavy ion collisions, chemical composition of hadrons is almost frozen
during expansion according to statistical model analyses.
Thus, each hadron acquires its chemical potential associated with
the approximated conserved number of the hadron below chemical freezeout temperature.
Again, the hadronic resonance gas model with finite chemical potential
\cite{Hirano:2002ds,Teaney:2002aj,Kolb:2002ve,Huovinen:2007xh,Qian:2007hj}
is needed to describe the space-time evolution of hadronic matter
in hydrodynamics.
Since the matter is already diluted due to expansion, one can instead use the hadronic
cascade model for a better dynamical description of hadronic matter.
This will be discussed in Sec.~\ref{sec:transport}.

\end{enumerate}

\subsection{Numerical aspects of relativistic ideal hydrodynamics}
\label{sec:num}

In this subsection, we review a numerical scheme to solve relativistic hydrodynamic
equations for perfect fluids.
We first discretize 
relativistic hydrodynamic equations in Cartesian coordinate  to solve numerically
in Sec.~\ref{sec:disc}.
We also discuss the conventional way to treat multi-dimensional
problem, namely, the operator splitting method.
After that, we introduce 
the piecewise parabolic method (PPM), \cite{Colella:1982ee}
which is known as a robust algorithm against strong shock waves,
to relativistic hydrodynamic equations in Sec.~\ref{sec:PPM}.
In practice, one needs to convert a set of numerical solutions
to physical quantities such as thermodynamic variables and flow velocity.
The procedure to obtain them is explained in Sec.~\ref{sec:thermo}.
Finally, we discuss hydrodynamic equations in relativistic coordinate
(proper time $\tau$ and space-time rapidity $\eta_{s}$)
in Sec.~\ref{sec:EqInBj}.

\subsubsection{Discretization and the operator splitting method}
\label{sec:disc}

In the Cartesian coordinate, hydrodynamic equations (\ref{eq:hydroeq}) and 
(\ref{eq:chargecons})
for a relativistic perfect fluid with one conserved charge 
(\textit{e.g.}, baryon charge)
can be written as
\begin{eqnarray}
\label{eq:hydroeqU}
\partial_t
\begin{pmatrix}
U_1 \\
U_2 \\
U_3 \\
U_4 \\
U_5 \\
\end{pmatrix}
+ \nabla \cdot
\begin{pmatrix}
U_1 \\
U_2 \\
U_3 \\
U_4 \\
U_5 \\
\end{pmatrix}
\bm{v} +
\begin{pmatrix}
\partial_x P \\
\partial_y P \\
\partial_z P\\
\nabla\cdot P\bm{v}  \\
0 \\
\end{pmatrix}
=0,
\end{eqnarray}
where
\begin{eqnarray}
\label{eq:defU}
\begin{pmatrix}
U_1 \\
U_2 \\
U_3 \\
U_4 \\
U_5 \\
\end{pmatrix}
& = & 
\begin{pmatrix}
\gamma^2(e+P)v_x \\
\gamma^2(e+P)v_y \\
\gamma^2(e+P)v_z \\
\gamma^2(e+P)-P \\
\gamma n_{B} \\
\end{pmatrix}.
\end{eqnarray}
These equations can be summarized as a continuity equation in the following form
\begin{eqnarray}
\partial_t U_J(t,\bm{x}) + \sum_{i=x,y,z}\partial_i F_{iJ} \left[U_J(t,\bm{x})\right] & = & 0, \quad
(J  =  1, \cdots, 5).
\label{eq:conteq}
\end{eqnarray}
For example, $F_{x1} = U_{1}v_{x} +P$, $F_{y1} = U_{1} v_{y}$, $F_{x2} = U_{2} v_{x}$ and so on.

One discretizes Eq.~(\ref{eq:conteq}) and write in a general form 
\begin{eqnarray}
[U_{J}]_{ijk}^{n+1} = [U_{J}]_{ijk}^{n} - \frac{\Delta t}{\Delta x}
\sum_{l=x,y,z} G_{lJ}\left[ [U_J]_{ijk}^n \right],
\end{eqnarray}
or more simply,
\begin{eqnarray}
U_{ijk}^{n+1} = U_{ijk}^{n} - \frac{\Delta t}{\Delta x} G_{l}\left[ U_{ijk}^n \right].
\end{eqnarray}
Here $n$ is a time step and $i$, $j$ and $k$ are fluid cell indices in $x$, $y$
and $z$ direction, respectively. $\Delta t$ and $\Delta x$ are mesh sizes in temporal
and spatial direction, respectively, which should obey
$C\leq 1$ in relativistic cases where $C = \Delta t /\Delta x$ is the Courant number
(Note that we have employed natural unit $c=1$).
We here assume isotropic lattice in three dimensions.
One can cope with these kinds of multi-dimensional equations
by employing the operator splitting method.
In this method, the operators are split into three sequential one-dimensional spatial steps:
\begin{eqnarray}
\tilde{U}_{ijk} & = & U_{ijk}^{n} - \frac{\Delta t}{\Delta x} G_x[U_{ijk}^{n}],\\
\hat{U}_{ijk} & = & \tilde{U}_{ijk} - \frac{\Delta t}{\Delta x} G_y[\tilde{U}_{ijk}],\\
U_{ijk}^{n+1} & = & \hat{U}_{ijk} - \frac{\Delta t}{\Delta x} G_z[\hat{U}_{ijk}].
\end{eqnarray}
To avoid numerical errors on spatial anisotropy, the above process is cyclically 
changed every time step.
Now the problem in three dimensional space reduces
to the one in one dimension. Hereafter in this subsection
we concentrate our discussion on solving
hydrodynamic equation in one-dimensional space.

\subsubsection{Piecewise parabolic method}
\label{sec:PPM}

As a robust numerical algorithm to solve hydrodynamic equations,
we review the piecewise parabolic method (PPM)\cite{Colella:1982ee} in this paper.
PPM was employed for the first time in
the physics of relativistic heavy ion collisions  in Ref.~\citen{Hirano:2000eu}
to solve 
relativistic hydrodynamic equations in Cartesian coodinate
and later applied to problems in relativistic $\tau$-$\eta_{s}$
coordinate in Ref.~\citen{Hirano:2001eu}.
PPM is categorized in the so-called Godunov's method.
Godunov made use of a Riemann's shock tube problem,
approximated a solution of interpolating two discontinuous
states to a constant one and obtained it
using conservation equations.
PPM is, so to speak, a higher order extension of the Godunov's approach.
As a consequence, PPM allows one to describe a hydrodynamic
response to steep profile very efficiently. For details, see Ref.~\citen{Colella:1982ee}.
For other algorithms such as SHASTA and rHLLE, see also Ref.~\citen{Schneider:1993gd}

For a given set of discrete values of
fields $\{U_{j}^{n}\}$ at time step $n$,
an interpolation function, $U_{j}(x)$, can be defined as
\begin{eqnarray}
\label{eq:Ujn}
U_{j}^{n} = \frac{1}{\Delta x} \int_{x_{j-1/2}}^{x_{j+1/2}} U_{j}(x) dx,
\end{eqnarray}
where $U_{j}(x)$ is assumed to be continuous and a parabolic function in $x_{j} -\Delta x/2 = x_{j-1/2} < x < x_{j+1/2} = x_{j} + \Delta x/2$.
Then one can parametrize $U_{j}(x)$ as
\begin{eqnarray}
\label{eq:Ux_param}
U_{j}(x) & = &U_{L,j} + \frac{x-x_{j-1/2}}{\Delta x}\left[\Delta U_{j} + 
U_{6,j}\left(1-\frac{x-x_{j-1/2}}{\Delta x} \right)\right],\\
\Delta U_{j} & = & U_{R,j}-U_{L,j},\\
U_{6,j} & = & 6\left[U_{j}^{n} -\frac{1}{2}\left(U_{R,j}+U_{L,j}\right)\right],
\end{eqnarray}
where $U_{j}(x_{j-1/2}) = U_{L,j}$ and
$U_{j}(x_{j+1/2}) = U_{R,j}$ to be determined as follows.

As default values at cell boundary, one puts
\begin{eqnarray}
\label{eq:URj}
U_{R,j} & = & \frac{7}{12}(U_{j}^{n} + U_{j+1}^{n})-\frac{1}{12}(U_{j+2}^{n}+U_{j-1}^{n}),\\
\label{eq:ULj}
U_{L,j} & = & \frac{7}{12}(U_{j-1}^{n} + U_{j}^{n})-\frac{1}{12}(U_{j+1}^{n}+U_{j}^{n}),
\end{eqnarray}
using discretized solutions $\{U_{j}^{n}\}$ at time step $n$.
These values are obtained as follows.
Assuming integral of $U(x)$ can be parameterized using a quartic function
(here $U(x)$ is rather defined globally at least in $x_{j-3/2}<x<x_{j+5/2}$),
\begin{eqnarray}
\mathcal{U}(x) & = & \int^{x} U(x') dx'\\
& = & a x^{4} +b x^{3} + c x^{2} + d x +e,
\end{eqnarray}
we calculate the value at the cell boundary $x=x_{j+1/2}$ as
\begin{eqnarray}
U(x_{j+1/2}) &=& \left. \frac{d\mathcal{U}}{dx}\right|_{x=x_{j+1/2}}
\end{eqnarray}
Only when we solve this problem, 
one can suppose $x_{j+1/2}=0$ and $\mathcal{U}(x=x_{j-3/2}) =0$
without loss of generality.
Then, $U(x_{j+1/2}) = d$.
We obtain the following coupled equations,
\begin{eqnarray}
\mathcal{U}(2\Delta x) & = & \int_{x_{j-3/2}}^{x_{j+3/2}}U(x) dx = \left(U_{j-1}+U_{j}+U_{j+1}+U_{j+2}\right)\Delta x,\\
\mathcal{U}(\Delta x) & = & (U_{j-1}+U_{j}+U_{j+1})\Delta x,\\
\mathcal{U}(0) & = & (U_{j-1}+U_{j})\Delta x,\\
\mathcal{U}(-\Delta x) & = & U_{j-1} \Delta x,\\
\mathcal{U}(-2\Delta x) & = & 0.
\end{eqnarray}
Solving these equations
with respect to $a, \cdots$ and $e$,
we finally obtain Eq.~(\ref{eq:URj}).

However, the value has to be reset in some cases. The interpolation function
$U_{j}(x)$ is imposed to be monotonic in $x_{j-1/2} < x < x_{j+1/2}$: $U_{j}(x)$ 
should take its value between $U_{R,j}$ and $U_{L,j}$.
In fact, $U_{j}(x)$ does not obey this condition either 
when $U_{j}^{n}$ is a local minimum or maximum
or when $U_{j}(x)$ has an extreme even though $U_{j}^{n}$ is between $U_{R,j}$ and $U_{L,j}$.
From Eq.~(\ref{eq:Ux_param}), this is the case when $\mid \Delta U_{j} \mid \geq \mid U_{6,j} \mid$.
The following replacement is made in these cases:
\begin{eqnarray}
U_{R,j}, U_{L,j} & \rightarrow & U_{j}^{n} \quad \mbox{if}  \quad(U_{R,j} - U_{j}^{n})(U_{j}^{n}- U_{L,j} )\leq 0,\\
U_{L, j} &\rightarrow &  3U_{j}^{n} - 2U_{R,j} \quad \mbox{if} \quad \Delta U_{j} U_{6,j} > (\Delta U_{j})^2,\\
U_{R, j} &\rightarrow &  3U_{j}^{n} - 2U_{L,j} \quad \mbox{if} \quad \Delta U_{j} U_{6,j} < -(\Delta U_{j})^2.
\end{eqnarray}

In the case of heavy ion collisions, fluids sometime can be surrounded by vacuum
in which $U_{j}^{n}$ vanishes.
In this case, both $U_{L,j}$ and $U_{R,j}$ are set to be zero.
When the default value calculated using Eq.~(\ref{eq:Ux_param}) becomes negative,
it is also set to be zero.\footnote{This might have been too strict for $U_{5}$
to be set
since it could be negative. However, if initial $U_{5}$ is positive, it keeps to be positive
in this fluid element.
Therefore there is no problem in ordinary cases.}

Using $U_{L,j}$ and $U_{R,j}$ determined above,
one calculates thermodynamic variables ($e_{L,R}$ and $P_{L,R}$), sound velocity
($c_{L, R}$) and flow velocity ($v_{L, R}$). We will discuss how to obtain
these variables from numerical solutions $U$ in the next subsection.
Average values of interpolation function
around the cell interfaces are
\begin{eqnarray}
\label{eq:Lbar}
\bar{U}_{L,j+1} & = &\frac{1}{\mid b_{r,j+1/2} \mid \Delta t}\int_{x_{j+1/2}}^{x_{j+1/2}+\mid b_{r,j+1/2} \mid \Delta t} U_{j+1}(x) dx,\\
\label{eq:Rbar}
\bar{U}_{R, j} & = & \frac{1}{\mid b_{l,j+1/2} \mid \Delta t}\int_{x_{j+1/2}-\mid b_{l,j+1/2} \mid \Delta t}^{x_{j+1/2}} U_{j}(x) dx
\end{eqnarray}
$b_r$ and $b_l$ are signal velocities
\begin{eqnarray}
b_{r,j+1/2} & = & \mbox{max}\left(0, \frac{v_{L,j+1} + c_{L,j+1}}{1+v_{L,j+1}c_{L,j+1}}, \frac{\bar{v}_{j+1/2} + \bar{c}_{j+1/2}}{1+\bar{v}_{j+1/2}\bar{c}_{j+1/2}}\right),\\
b_{l,j+1/2} & = & \mbox{min}\left(0, \frac{v_{R,j} - c_{R,j}}{1-v_{R,j}c_{R,j}}, \frac{\bar{v}_{j+1/2} + \bar{c}_{j+1/2}}{1+\bar{v}_{j+1/2}\bar{c}_{j+1/2}}\right),\\
\bar{v}_{j+1/2} & = & \frac{1}{2}(v_{R,j}+v_{L,j+1}),\\
\bar{c}_{j+1/2} & = & \frac{1}{2}(c_{R,j}+c_{L,j+1}).
\end{eqnarray}
Notice that $b_{l,j+1/2}\le 0$.
When a fluid cell is located next to vacuum, \textit{e.g.}, $(U_{4})_{j} \neq 0$ and 
$(U_{4})_{j+1} = 0$, we set $b_{r} = 1$.
Please notice that the subscripts $R$ and $L$
represent, respectively, the values of right side and left side \textit{at each cell},
but that the subscripts $r$ and $l$ denote, respectively,
the values of right side and left side \textit{at each cell boundary}.
Inserting Eq.~(\ref{eq:Ux_param}) into Eqs.~(\ref{eq:Lbar}) and (\ref{eq:Rbar}),
we obtain
\begin{eqnarray}
\bar{U}_{L,j+1} & = & U_{L,j+1} + \frac{b_{r, j+1/2}\Delta t}{2\Delta x}
\left[\Delta U_{j+1} + \left(1-\frac{2b_{r,j+1/2} \Delta t}{3\Delta x} \right)U_{6,j+1} \right],\\
\bar{U}_{R,j} & = & U_{R,j} - \frac{b_{l, j+1/2}\Delta t}{2\Delta x}
\left[\Delta U_{j} - \left(1-\frac{2b_{l,j+1/2} \Delta t}{3\Delta x} \right)U_{6,j} \right].
\end{eqnarray}
By using these average values above, 
we solve the Riemann problem at each cell boundary.
The solution becomes complicated in general.
Thus, in the Godunov-type algorithm,
one approximates the solution to a constant value $U_{lr}$ which fulfills
the conservation law.
One rewrites the hydrodynamic equations in their integral form:
\begin{eqnarray}
&& \int_{x_{j}}^{x_{j+1/2}} [U(x, t_{n}+\Delta t/2) - U(x,t_{n})]dx \nonumber\\
 & = & -\int_{t_{n}}^{t_{n}+\Delta t/2}
\left[F(U(x_{j+1/2}, t)) - F(U(x_{j},t)) \right]dt
\end{eqnarray}
The integration can be easily done by assuming
$U(x_{j}) = \bar{U}_{R,j}$ = const.~in $t_{n} < t < t_{n}+\Delta t/2$.
The result becomes
\begin{eqnarray}
\label{eq:barUr}
(\bar{U}_{R,j}-U_{lr})b_{l,j+1/2}\frac{\Delta t}{2} & = & -\left[F(U_{lr})-F(\bar{U}_{R,j}) \right]\frac{\Delta t}{2}.
\end{eqnarray}
One also integrates the hydrodynamic equations over $x_{j+1/2}<x<x_{j+1}$
\begin{eqnarray}
&&\int_{x_{j+1/2}}^{x_{j+1}} [U(x, t_{n}+\Delta t/2) - U(x,t_{n})]dx \nonumber \\
 & = & -\int_{t_{n}}^{t_{n}+\Delta t/2}
\left[F(U(x_{j+1}, t)) - F(U(x_{j+1/2},t)) \right]dt
\end{eqnarray}
and obtains
\begin{eqnarray}
\label{eq:barUl}
(U_{lr}-\bar{U}_{L,j+1})b_{r,j+1/2}\frac{\Delta t}{2} & = & -\left[F(\bar{U}_{L,j+1})-F(U_{lr}) \right]\frac{\Delta t}{2}.
\end{eqnarray}
From Eqs.~(\ref{eq:barUr}) and (\ref{eq:barUl}), one obtains $U_{lr}$ and $F(U_{lr})$
\begin{eqnarray}
U_{lr} & = & \frac{F(\bar{U}_{R,j})-F(\bar{U}_{L,j+1})-b_{l,j+1/2}\bar{U}_{R,j}+b_{r,j+1/2}\bar{U}_{L,j+1}}{b_{r,j+1/2}-b_{l,j+1/2}}\\
F(U_{lr}) & = & F_{j+1/2}\nonumber \\
 & = & \frac{b_{r,j+1/2} F(\bar{U}_{R,j}) - b_{l,j+1/2} F(\bar{U}_{L,j+1}) + b_{r,j+1/2} b_{l,j+1/2}(\bar{U}_{L,j+1}-\bar{U}_{R,j})}{b_{r,j+1/2}-b_{l,j+1/2}}\nonumber \\
\end{eqnarray}
Finally, the solution at the next time step leads to
\begin{eqnarray}
U_{j}^{n+1} & = & U_{j}^{n} -\frac{\Delta t}{\Delta x} (F_{j+1/2}-F_{j-1/2})
\end{eqnarray}
The key difference between the conventional Godunov scheme and the present method
is using the average values $\bar{U}_{R,j}$ and $\bar{U}_{L,j+1}$ instead of
$U_{j}$ and $U_{j+1}$ to gain  accuracy of numerical solutions
since numerical fluxes using $U_{j}$ or $U_{j+1}$
are often overestimated, in particular, in the case of steep profiles.

\subsubsection{Thermodynamic variables and flow velocity}
\label{sec:thermo}

We transform from thermodynamic variables and flow velocity
to variables to solve the relativistic hydrodynamic equations numerically in Eq.~(\ref{eq:defU}).
Hence we need to transform back to thermodynamic variables from numerical
solutions $U_{J}$.
From Eq.~(\ref{eq:defU}),
\begin{eqnarray}
\label{eq:eqv}
v = \mid \bm{v} \mid & = & \frac{\mid \bm{U} \mid }{U_{4}+P(e, n_{B})},\\  
\label{eq:e}
e & = & U_{4} - \bm{U}\cdot \bm{v},\\
\label{eq:nb}
n_{B}  & = & U_{5} \sqrt{1-v^2},\\
\bm{U} & = &(U_{1}, U_{2}, U_{3}).
\end{eqnarray}
For a given equation of state, $P=P(e,n_{B})$, and numerical solutions, $U_{J}$,
Eq.~(\ref{eq:eqv}) with Eqs.~(\ref{eq:e}) and (\ref{eq:nb}) 
becomes a non-linear equation with respect to $v$. Therefore,
one has to solve it numerically using, \textit{e.g.}, the bi-section method.
Once the solution $v$ is obtained from Eq.~(\ref{eq:eqv}),
it is easy to obtain energy density and
baryon density from Eqs.~(\ref{eq:e}) and (\ref{eq:nb}) and, consequently, 
pressure from the equation of state.

\subsubsection{Hydrodynamic equations in relativistic coordinate}
\label{sec:EqInBj}

It is more appropriate to write down relativistic hydrodynamic equations
in an expanding coordinate in the case of relativistic heavy ion collisions:

\begin{eqnarray}
\label{eq:hydroeqUrela}
\partial_{\tau}
\begin{pmatrix}
U_1 \\
U_2 \\
U_3 \\
U_4 \\
U_5 \\
\end{pmatrix}
+ \nabla \cdot
\begin{pmatrix}
U_1 \\
U_2 \\
U_3 \\
U_4 \\
U_5 \\
\end{pmatrix}
\tilde{\bm{v}} +
\begin{pmatrix}
\tau \partial_x P \\
\tau \partial_y P \\
\partial_{\eta_{s}} P\\
\tau \nabla \cdot P\tilde{\bm{v}}  \\
0 \\
\end{pmatrix}
+
\begin{pmatrix}
0 \\
0 \\
U_{3}/\tau \\
U_{4} \tilde{v}_{\eta_{s}}^{2}/\tau + P(1+\tilde{v}_{\eta_{s}}^{2})  \\
0 \\
\end{pmatrix}
=0,
\end{eqnarray}
where
\begin{eqnarray}
\label{eq:defUrela}
\begin{pmatrix}
U_1 \\
U_2 \\
U_3 \\
U_4 \\
U_5 \\
\end{pmatrix}
& = & 
\begin{pmatrix}
\tau \tilde{\gamma}^2(e+P)\tilde{v}_x \\
\tau \tilde{\gamma}^2(e+P)\tilde{v}_y \\
\tau \tilde{\gamma}^2(e+P)\tilde{v}_z \\
\tau \tilde{\gamma}^2(e+P)-\tau P \\
\tau \tilde{\gamma} n_{B} \\
\end{pmatrix}.
\end{eqnarray}
where proper time $\tau = \sqrt{t^2-z^2}$, space-time rapidity
$\eta_{s} = (1/2)\ln[(t+z)/(t-z)]$ and 
$\nabla = (\partial_{x}, \partial_{y}, \partial_{\eta_{s}}/\tau)$.
Flow velocities and fluid rapidity in this coordinate are, respectively,
\begin{eqnarray}
\tilde{v}_{x} & = & \frac{\cosh Y_{f}}{\cosh(Y_{f} - \eta_{s})} v_{x} ,\\
\tilde{v}_{y} & = & \frac{\cosh Y_{f}}{\cosh(Y_{f} - \eta_{s})} v_{y} ,\\
\tilde{v}_{\eta_{s}} & = & \tanh(Y_{f}-\eta_{s}),\\
\tilde{\gamma} & = & \frac{1}{\sqrt{1-\tilde{v}_{x}^{2}-\tilde{v}_{y}^{2}-\tilde{v}_{\eta_{s}}^{2}}}.
\end{eqnarray}
where $Y_{f}  =  \frac{1}{2}\ln\left[(1+v_{z})/(1-v_{z})\right]$.

Equation (\ref{eq:hydroeqUrela})
is quite similar to Eq.~(\ref{eq:hydroeqU}) except for 
existence of the last term in the left hand side
which is a source term due to expanding coordinate.
Therefore, one can utilize the same PPM algorithm by adding source terms
in the cycle of operator splitting.

One can obtain thermodynamic variables and flow velocities 
from the following relations:
\begin{eqnarray}
\label{eq:V}
\tilde{v} = \mid \tilde{\bm{v}} \mid & = &\frac{\mid \bm{U} \mid/\tau}{U_{4}/\tau + p(e, n_B)},\\
\label{eq:E}
e & = & \frac{U_4}{\tau}-\frac{\mid \bm{U} \mid }{\tau}\cdot \mid \tilde{\bm{v}} \mid , \\
\label{eq:NB}
n_B & = &\frac{U_5}{\tau}\sqrt{1-\mid \tilde{\bm{v}} \mid^2}.
\end{eqnarray}
Inserting Eqs.~(\ref{eq:E}) and (\ref{eq:NB}) into Eq.~(\ref{eq:V}), we first
solve an implicit equation for $\mid \tilde{\bm{v}} \mid$ numerically
and then obtain $e$, $n_B$ and $P=P(e, n_B)$. From numerical
solutions $U_{J}$,
we also obtain each component of $\tilde{\bm{v}}$:
\begin{eqnarray}
\tilde{\bm{v}} = \frac{1}{U_{4} + \tau P}(U_1, U_2, U_3).
\end{eqnarray}
Velocities in the Cartesian coordinate are obtained from $\tilde{\bm{v}}$
\begin{eqnarray}
v_x & = & \frac{\cosh (Y_{f}-\eta_{s})}{\cosh Y_{f}}\tilde{v}_{x}, \\
v_y & = & \frac{\cosh (Y_{f}-\eta_{s})}{\cosh Y_{f}}\tilde{v}_{y}, \\
v_z & = & \frac{\tilde{v}_{\eta_{s}} \cosh \eta_{s} + \sinh \eta_{s}}{\cosh \eta_{s} + \tilde{v}_{\eta_{s}} \sinh \eta_{s}} = \tanh Y_{f},\\
\gamma & = & \frac{\cosh Y_{f}}{\cosh(Y_{f}-\eta_{s})}\tilde{\gamma}, 
\end{eqnarray}
where $Y_{f} = \tanh^{-1} \tilde{v}_{\eta_{s}} + \eta_{s}$.

\section{Initial Conditions}
\label{sec:initialcondition}

We need to specify  initial conditions for hydrodynamic simulations.
In principle, we must understand the initial particle production
and subsequent non-equilibrium evolution of the system toward thermal
state to obtain the initial condition.
However, at the moment, we do not have complete understanding
at early stages of high energy nuclear collisions.
Here, we make a simple assumption that produced particles
after the collision of two nuclei can be used to obtain
initial entropy distribution.
We will come back to this issue on isotropization at an early stage
of collisions from a viewpoint of non-Abelian
plasma instability in Sec.~\ref{sec:nonabelian}.


The $k_T$-factorization formulation is widely used to compute the inclusive
cross section for produced gluons~\cite{KLN,Gelis:2006tb,Albacete:2007sm,
Levin:2011hr,Tribedy:2010ab}
in hadronic collisions.
In order to study gluon production in nucleus-nucleus collision,
we shall use the Monte-Carlo implementation of $k_T$-factorization formulation
or Glauber model (MC-Glauber) 
\cite{Miller:2007ri,Broniowski:2007nz,Alver:2008aq}
in which fluctuations of the position of nucleons inside a nucleus
are taken into account.
Thus we can study nucleus-nucleus collisions
on an event-by-event basis.

We first sample the positions of nucleons inside a
nucleus according to a nuclear density distribution (\textit{e.g.}, Woods-Saxon function)
for two colliding nuclei,
and shift them by a randomly-chosen impact parameter $b$
with probability $b\,db$ for an event.
A nucleon-nucleon collision takes place if their distance $d$ in the
transverse plane orthogonal to the beam axis fulfills the condition
\begin{equation}
  d \leq \sqrt{\frac{\sigma_\text{in}}{\pi}}\ ,
\end{equation}
where $\sigma_\text{in}$ denotes the inelastic nucleon-nucleon cross section.
Incident energy dependent total $pp$ cross section is
parameterized by Particle Data Group~\cite{pdg1996}.
Elastic cross section is computed
using PYTHIA parametrization~\cite{Schuler:1993td,Sjostrand:2006za}.
The following values are obtained; 
$\sigma_\text{in}=39.53, 41.94$ and 61.36 mb
at $\sqrt{s}=130, 200$ and 2760 GeV, respectively.
In this way, we obtain the number of binary collisions and that of participants for each event. 
It should be noticed that the standard Woods-Saxon parameters
shown in, \textit{e.g.}, Ref.~\citen{DeJager:1987qc}
cannot be directly used to distribute nucleons inside a nucleus
because of the finite interaction range in our approach.
We need to modify nuclear density parameters
so that a convolution of nucleon profiles leads to the measured 
Woods-Saxon profile~\cite{Hirano:2009ah}.

Next, we compute particle production at each grid in the transverse plane.
In the MC-Glauber approach, we assume that the initial entropy profile
in the transverse plane is proportional to a linear combination of
the number density of participants and that of binary collisions:
\begin{equation}
s_0(\bm{x}_\perp) 
 \equiv 
 \left. \frac{dS}{\tau_0 dx dy d\eta_s} \right|_{\eta_s = 0}
 =  \frac{C}{\tau_0}\left(\frac{1-\alpha}{2}
    \rho_{\mathrm{part}}(\bm{r}_\perp)
  + \alpha\, \rho_{\mathrm{coll}}(\bm{r}_{\perp})
   \right),
\end{equation}
where $\tau_0=0.6$ fm/$c$ is a typical initial time for the hydrodynamical
simulation.
Parameters $C=19.8$ and $\alpha=0.14$  have been fixed through
comparison with the centrality dependence of multiplicity data
in Au+Au collisions at RHIC \cite{PHOBOS_Nch} by pure hydrodynamic
calculations with temperature $T_{\mathrm{dec}}=100$\,MeV.\cite{Hirano:2009ah}
At the LHC energy, $C=41.4$ and $\alpha = 0.08$ are chosen \cite{Hirano:2010je}
so that we reproduce the ALICE data on centrality dependence 
of multiplicity in Pb+Pb collisions at $\sqrt{s_{NN}}=2.76$ TeV.\cite{ALICEdNdeta,Aamodt:2010ft}

The participant density $\rho_\text{part}(\bm{r}_\perp)$ at each grid point
is the sum of participants density $\rho_A(\bm{r}_\perp)$
from nucleus $A$ and $\rho_B(\bm{r}_\perp)$ from nucleus $B$, which
are computed by counting the number of wounded nucleons  $N_{A,w}$
and $N_{B,w}$ for nucleus $A$ and $B$
within a tube extending in the beam direction with the radius
$r=\sqrt{\sigma_\text{in}/\pi}$:
\begin{equation}
\rho_\text{part}(\bm{r}_\perp) 
=\rho_\text{A}(\bm{r}_\perp) + \rho_\text{B}(\bm{r}_\perp) 
= \frac{N_w}{\sigma_\text{in}}
\end{equation}
Similarly, the number of binary collision density at each grid is obtained by
counting the number of binary collision $N_{\mathrm{coll}}$ 
with the area $\sigma_\text{in}$, where the transverse position
of binary collision is assumed to be the average transverse coordinate
between two colliding nucleons
\begin{equation}
   \rho_{\mathrm{coll}}(\bm{r}_{\perp})
   =\frac{N_{\mathrm{coll}}}{\sigma_\text{in}}\ .
\end{equation}
which may be also obtained by the expression
$\rho_A(\bm{r}_\perp)\rho_B(\bm{r}_\perp)\sigma_\text{in}$.

In the Monte-Carlo KLN (MC-KLN) model~\cite{MCKLN},
the number distribution of gluon production
at each transverse grid is given by
the $k_T$-factorization formula~\cite{KLN}
\begin{eqnarray}
  \frac{dN_g}{d^2 r_{\perp}dy} & = &\kappa
   \frac{4N_c}{N_c^2-1}
    \int
    \frac{d^2p_\perp}{p^2_\perp}
      \int \frac{d^2k_\perp}{4} \;\alpha_s(Q^2)\nonumber \\
       &\times &   \phi_A(x_1,(\bm{p}_\perp + \bm{k}_\perp)^2/4)\;
       \phi_B(x_2,(\bm{p}_\perp{-}\bm{k}_\perp)^2/4)~,
      \label{eq:ktfac}
\end{eqnarray}
with $N_c=3$ the number of colors.
Here, $p_\perp$ and $y$ denote the
transverse momentum and the rapidity of the produced gluons, respectively. 
The light-cone momentum fractions of the colliding gluon
ladders are then given by $x_{1,2} = p_\perp\exp(\pm y)/\sqrt{s_{NN}}$,
where $\sqrt{s_{NN}}$ denotes the center of mass energy.
Running coupling $\alpha_s(Q^2)$ is evaluated at the scale
$Q^2=\max( (\bm{p}_\perp -\bm{k}_\perp)^2/4,(\bm{p}_\perp
+\bm{k}_\perp)^2/4)$.
An overall normalization factor $\kappa$ is so chosen that
the multiplicity data in Au+Au collisions at RHIC are fitted
in most central collisions.
In the MC-KLN model, saturation momentum
is parameterized by assuming that the saturation momentum square is 2 GeV$^2$
at $x=0.01$
in Au+Au collisions at $b=0$ fm at RHIC where $\rho_\text{part}=3.06$
fm$^{-2}$ \cite{KLN}:
\begin{equation}
Q_{s,A}^2 (x; \bm{r}_\perp)  =  2\ \text{GeV}^2
\frac{\rho_{A}(\bm{x}_\perp)}{1.53\ \text{fm}^{-2}}
\left(\frac{0.01}{x}\right)^{\lambda} \ ,
\label{eq:qs2}
\end{equation}
where $\lambda$ is a free parameter which is expected
to have the range of $0.2<\lambda<0.3$ from
Hadron Electron Ring Accelerator (HERA) global analysis for
$x<0.01$~\cite{HERA}.
In MC-KLN, we assume the gluon distribution
function as
\begin{equation}
\label{eq:uninteg}
  \phi_{A}(x,k_\perp^2;\bm{r}_\perp)\sim
    \frac{1}{\alpha_s(Q^2_{s,A})}\frac{Q_{s,A}^2}
       {{\rm max}(Q_{s,A}^2,k_\perp^2)}~,
\end{equation}

We assume that initial conditions of hydrodynamical simulations
are obtained by identifying the gluons' momentum rapidity $y$ with
space-time rapidity $\eta_s$
\begin{equation}
s_0(\bm{x}_\perp) 
\propto \frac{dN}{\tau_0 d\bm{x}_\perp d\eta_s}
\end{equation}

Note here that recently gluon distribution function $\phi$
obtained from numerical results of
the running coupling Balitsky-Kovchegov (rcBK) evolution
equation~\cite{Albacete:2007yr,Albacete:2009fh,Albacete:2010sy}
\begin{equation}
\frac{\partial\mathcal{N}(r,x)}{\partial y}
= \int d^2 r_1 \ K(r,r_1,r_2) \left[\mathcal{N}(r_1,y)+\mathcal{N}(r_2,y)
  -\mathcal{N}(r,y)-\mathcal{N}(r_1,y)\,\mathcal{N}(r_2,y)\right]
\end{equation}
is employed in the more sophisticate model called the MCrcBK
model~\cite{mckt},
where $r_2 = r - r_1$.
$\phi$ is obtained from the Fourier transform of
the numerical results of the rcBK evolution equation:
\begin{equation}
\phi_{A,B}(k,x,\bm{r}_\perp)=\frac{C_F}{\alpha_s(k)\,(2\pi)^3}\int d^2 \bm{r}\
e^{-i \bm{k} \cdot \bm{r}}\,\nabla^2_{\bm{r}}\,
(1-\mathcal{N}(r,y))^2
\label{phi}
\end{equation}
where $C_F=(N_c^2-1)/2N_c$, $y=\log(x_0/x)$ and  $x_0=0.01$, 
In MC-KLN model, $x$ dependence is determined by Eq.~(\ref{eq:qs2}),
but in rcBK, $x$ dependence can be obtained from the equation.
Therefore, we expect that MCrcBK model has more predictive power than
MC-KLN model. See Refs.~\citen{Albacete:2007sm,Albacete:2010bs,Tribedy:2010ab,
mckt,JalilianMarian:2011dt}
for the predictions of hadron productions
in nuclear collisions using rcBK solutiuons.

One can use a Gaussian shape
for nucleons~\cite{Albacete:2011fw,Nara:2011ee,Heinz:2011mh} in the
Monte-Carlo method.
This smooth density profile for a nucleon may be significant
for the simulation of event-by-event viscous hydrodynamics~\cite{Heinz:2011mh}.
In this case the thickness function for a nucleon is given by
\begin{equation}
T_p(r) = \frac{1}{2\pi B}\exp[-r^2/(2B)]\ .
\end{equation}
The probability of  a nucleon-nucleon collision $P(b)$ at an impact parameter $b$
is then taken to be
\begin{equation}
P(b) = 1 - \exp[-k T_{pp}(b)],\qquad T_{pp}(b) = \int d^2s \, T_p(s)\,
T_p(s-b)\ .
\end{equation}
where (perturbatively) $k$ corresponds to the product of gluon-gluon
cross section
and gluon density squared.
We fix $k$ so that integral with respect to the impact parameter
becomes the nucleon-nucleon inelastic cross section $\sigma_{NN}$ at
the given energy:
\begin{equation}
\sigma_{NN}(\sqrt{s}) = \int d^2b \left(
   1-\exp[-k(\sqrt{s}) \, T_{pp}(b)]
    \right) ~.
\end{equation}
Note that $P(b)$ broadens with increasing energy, even as the size $B$ of the
hard valence partons is fixed.

We finally note that fluctuations from gluon production
can be included in the model~\cite{Tribedy:2010ab,Dumitru:2012yr}
(for MC-Glauber model see Ref.~\citen{Qin:2010pf}),
which is very large contribution to the initial higher order
anisotropies.

\section{Energy Loss of Jets inside the Expanding Fluid}
\label{sec:energyloss}

Jet quenching is one of the promising tools to diagnose dense 
matter created in relativistic heavy ion collisions.
Energetic partons created in hard scatterings
are subject to traverse dense medium.
Through interaction between these partons
and medium, various consequences are predicted to take places: 
suppression of yields for high $p_{T}$ hadrons, 
energy imbalance of a jet pair and so on.
In this section we overview mainly how to calculate the amount of energy loss
in hydrodynamic backgrounds. 
For details about existing energy loss formalisms 
such as BDMPS-Z,\cite{BDMPSZ} GLV,\cite{GLV} ASW,\cite{ASW} HT\cite{HT}
 and AMY\cite{AMY}
and comparison among them, 
see \textit{e.g.} Ref.~\citen{Armesto:2011ht}. 

First attempt to combine energy loss calculation with hydrodynamic simulations
was made in Ref.~\citen{Gyulassy:2001kr}.
After that, the hydro + jet model
based on mini-jet production from PYTHIA \cite{pythia} and
its propagation in
full three dimensional ideal hydrodynamic
backgrounds \cite{Hirano:2001eu,Hirano:2002ds}
has been developed and
systematic analysis has been performed
including $p_{T}$ spectra, di-hadron correlation
functions, interplay between soft and hard components and
suppression in forward rapidity regions.
\cite{Hirano:2002sc,Hirano:2003hq,hydrojet}

In most of energy loss formalisms, the amount of energy loss
largely depends on the inverse of mean free path $\lambda^{-1}$,
and, in turn, on the medium parton density $\rho = (\sigma \lambda)^{-1}$
from the kinetic theory, where $\sigma$ is the cross section between
an emitted gluon and a parton in the medium.
For example, jet quenching (transport) parameter can be
 $\hat{q} \approx \mu^2/\lambda$,
where $\mu$ is typical transverse momentum transfer suffered from the medium. 
When an ideal gas of massless quarks and gluons is adopted for the QGP equation of state,
medium parton density is calculated from
thermodynamic variables, $\rho \propto e^{3/4} \propto T^{3}$.\cite{Baier:2002tc}
In this way, one can utilize hydrodynamic outputs along a trajectory of a jet
to quantify energy loss in relativistic heavy ion collisions.
It should be noted that the above relation is valid only for the ideal gas equation of state.
Since the medium is strongly coupled according to hydrodynamic analyses of collective 
anisotropic flow,
non-perturbative definition of the energy loss is demanded.\cite{Majumder:2012sh}

Initial transverse positions of jets at an impact parameter
$\bm{b}$ are determined randomly according to the probability distribution
specified by the number of binary collision
distribution, 
\begin{eqnarray}
P(\bm{r}_{\perp}, \bm{b}) & \propto & T_{A}(\bm{r}_{\perp} + \bm{b}/2)T_{A}(\bm{r}_{\perp} -\bm{b}/2)
\end{eqnarray}
Initial longitudinal position of a parton is approximated
by the boost invariant distribution: $\eta_{s} = y$, where
$y = (1/2) \ln[(E + p_{z})/(E -p_{z})]$
is the rapidity of a parton.
Jets are freely propagated up to the initial time $\tau_{0}$
of hydrodynamic simulations by neglecting the possible
interactions in the pre-thermalization stages.
Jets are assumed to travel with straight line trajectory in a time
step:
\begin{eqnarray}
\Delta r_{i} & = & \frac{p_{i}}{ m_{T} \cosh(Y - \eta_{s})} \Delta \tau, \enskip (i = x, y),\\
\Delta \eta_{s} & = & \frac{1}{\tau} \tanh(Y -\eta_{s})\Delta \tau, 
\end{eqnarray}
where $m_{T} = \sqrt{m^{2} + p_{T}^{2}}$
is a transverse mass.
We obtain the total amount of energy loss for a sample jet
\begin{eqnarray}
\Delta E \propto \int_{\tau_0}^{\infty}d\tau \rho(\bm{x}(\tau), \tau)(\tau-\tau_{0})
\ln\left(\frac{2E_{0}}{\mu^2 L} \right)
\end{eqnarray}
when we employ the approximated first order formula in the opacity expansion.\cite{GLV}
Here $\tau_0$ is the initial time of hydrodynamic simulations, $E_0$ is the initial
energy of a jet which is Lorentz boosted by the four flow velocity
as $p_{0}^{\mu} u_\mu$. The formula roughly gives $L^2$ dependence of energy
loss in the case of a static medium, 
which is manifestation of the Landau-Pomeranchuk-Migdal effect.

It would be interesting to see a response of the medium to parton energy loss.
This could be neglected at the RHIC energy where initial energy of jets are not so large.
On the other hand, jets with a few hundred GeV
can be produced at the LHC energy.
If the lost energy is quickly thermalized,
one can solve hydrodynamic equations with a source term from energy loss.
\begin{eqnarray}
\partial_\mu T^{\mu \nu} (x) & = & J^{\nu}(x),\\
J^{0} & = & J^{1} = -\frac{dp^{0}}{dt}\delta(x-x_0-t)\delta(y-y_0)\delta(z-z_0),\\
J^{2} & = & J^{3} = 0,
\end{eqnarray}
where initial position of a jet is ($x_0$, $y_0$, $z_0$)
and the jet is supposed to traverse in a straight trajectory in the $x$ direction.  
These equations can be solved numerically using the algorithm mentioned
in Sec.~\ref{sec:num}.

\section{Transport Model}
\label{sec:transport}

Relativistic transport models have been successfully used to
simulate from low to high energy nuclear collisions.
A great advantage of transport models is that one can study
space time evolution of system
even if  it is not equilibrated.
Transport models include, for example,
intra nuclear cascade models~\cite{Cugnon:1980zz,Cugnon:1980rb,Kodama:1983yk},
Boltzmann-Uehling-Uhlenberck (BUU) models~\cite{Bertsch:1988ik},
RQMD~\cite{rqmd1,rqmd2}, QGSM~\cite{qgsm}, ARC~\cite{arc},
ART~\cite{Li:1995pra}, UrQMD~\cite{urqmd}, JAM~\cite{JAM,Isse:2005nk},
HSD~\cite{Cassing:1999es},
and  GiBUU~\cite{Buss:2011mx}.
The partonic phase in the early stages of heavy ion collisions
has been studied by the parton cascade models; VNI~\cite{vni},
VNI/BMS~\cite{Bass:2002fh,Renk:2005yg},
ZPC~\cite{Zhang:1997ej}, MPC~\cite{Molnar:2000jh},
 GROMIT~\cite{Cheng:2001dz}
and BAMPS~\cite{Xu:2004mz}.
Partonic cascade in which only the elastic scattering
is included does not account
for the observed elliptic flow for typical gluon-gluon pQCD
cross section of $\sigma_{gg} \sim 3$ mb~\cite{Molnar:2001ux}.
Inelastic process such as $2\to 3$ scattering has been
shown to be very important for the energy loss of the
high energy jets as well as the thermalization of the system.~
\cite{vni,chem,SMHWong,Nara:2001zu,Xu:2004mz,Renk:2005yg}
The transport models
such as the multi-phase transport model (AMPT)~\cite{Lin:2004en}
and the Paton-Hadron-String Dynamics (PHDS)~\cite{Cassing:2009vt}
include the dynamics of both partonic as well as
hadronic phase.
In the following, we shall review mainly the transport model JAM~\cite{JAM}.

\subsection{Scattering algorithms}
The geometrical interpretation of the cross section 
for collisions of particles
is employed by various models,
\textit{i.e.}, particles scatter when their
closest distance is smaller than $\sqrt{\sigma/\pi}$, where
$\sigma$ is a total cross section.
The particles are propagated along classical trajectories until
they scatter or decay. The scattering of two particles is determined
by the method of closest distance approach; two particles
scatter, if the impact parameter (closest distance) $b_{\mathrm{rel}}$
for a pair of particles becomes less than the
interaction range specified by the cross section:
\begin{equation}
  b_{\mathrm{rel}} \leq \sqrt{\frac{\sigma(\sqrt{s})}{\pi}}
\end{equation}
where $\sigma(\sqrt{s})$ is the total cross section for the pair
at the c.m.~energy $\sqrt{s}$ which depends on the incoming
particle pair.
This collision method has been widely used to simulate high energy
nucleus-nucleus collisions.
Because of a geometrical interpretation of the cross section,
this method violates causality, and time ordering of the collisions
in general differs from one frame to another.
Those problems have been studied by several
authors~\cite{Kodama:1983yk,Zhang:1996gb}

Our approach is motivated by Kodama's work~\cite{Kodama:1983yk,Wolf:1990ur}
in which Lorentz invariant expressions are obtained by considering
the rest frame of one of colliding particles.
The closest distance $b_{\mathrm{rel}}$ is defined by the distance in their common
c.m. frame. Suppose that the coordinates and momenta
of two colliding particles are denoted by
$x_1 = (t_1, \bm{x}_1)$,
$x_2 = (t_2, \bm{x}_2)$,
$p_1 = (E_1, \bm{p}_1)$
and $p_2 = (E_2, \bm{p}_2)$.
The trajectories of particles are 
\begin{align}
 \bm{x}^*_1(t^*) &= \bm{x}^*_1(t^*_1) + \bm{v}_1^* ( t^*- t^*_1),\nonumber\\
 \bm{x}^*_2(t^*) &= \bm{x}^*_2(t^*_2) + \bm{v}_2^* ( t^*- t^*_2),
 \label{eq:motion}
\end{align}
where asterisks represent quantities in the two-body c.m. frame.
The time of closest approach $t^*_c$ may be obtained by the 
condition that the relative separation becomes perpendicular to
the relative momentum, 
$(\bm{x}^*_1-\bm{x}^*_2)\cdot (\bm{p}^*_1-\bm{p}^*_2)=0$,
\begin{align}
 t^*_c - t^*_1 
 &=-\frac{\bm{v}^*\cdot(\bm{x}^* - \bm{v}_2^* (t_1^*-t_2^*))}
                        {\bm{v}^{*2}}
 =-\frac{\bm{v}^*\cdot(\bm{x}_1^*(t^*_1) - \bm{x}_2^*(t^*_1))}
                        {\bm{v}^{*2}}, \\
 t^*_c - t^*_2
 &=-\frac{\bm{v}^*\cdot(\bm{x}^* + \bm{v}_1^* (t_2^*-t_1^*))}
                        {\bm{v}^{*2}}
 =-\frac{\bm{v}^*\cdot(\bm{x}_1^*(t^*_2) - \bm{x}_2^*(t^*_2))}
                        {\bm{v}^{*2}},
\label{eq:time}
\end{align}
where $\bm{x}^* = \bm{x}_1^*(t_1^*) - \bm{x}_2^*(t_2^*)$,
$t^* = t_1^* - t_2^*$,
and $\bm{v}^*=\bm{v}_1^* - \bm{v}_2^*$.
Then the closest distance $b_{\mathrm{rel}}$ is expressed as
\begin{equation}
 b^2_{\mathrm{rel}} = \bm{x}^{*2}
 - \frac{(\bm{x}^*\cdot \bm{v}^*)^2}{\bm{v}^{*2}}
\end{equation}
One can express $b_{\mathrm{rel}}$ in terms of Lorentz invariant scalars
\begin{equation}
 b_{\mathrm{rel}}^2 = -x^2 + \frac{(P\cdot x)^2}{P^2} + \frac{(q\cdot x)^2}{q^2}
\end{equation}
by using the following variables
\begin{align}
 x &= x_1 - x_2 = (t_1 - t_2, \bm{x}_1 - \bm{x}_2), \\
 p &= p_1 - p_2 = (E_1 - E_2, \bm{p}_1 - \bm{p}_2), \\
 P &= p_1 + p_2 = (E_1 + E_2, \bm{p}_1 + \bm{p}_2),\\
 q &= p -\frac{P\cdot p}{P^2}P,
\end{align}
where we have used the Lorentz invariant expressions for
$-(\bm{x}^*\cdot \bm{p}^*)^2/\bm{p}^{*2} = (q\cdot x)^2/q^2$, and transverse
square distance $-x^2+(P\cdot x)^2/P^2$ corresponds to the squared distance
$\bm{x}^{*2}$ in the two-body c.m. frame.  The following expression for impact
parameter in terms of the original momenta will be useful \begin{equation}
b_\text{rel}^2 = -x^2 - \frac{p_1^2(p_2\cdot x)^2 + p_2^2(p_1\cdot x)^2
-2(p_1\cdot p_2)(p_1\cdot x)(p_2\cdot x)} {(p_1\cdot p_2)^2-p_1^2p_2^2}\ .
\end{equation}

One can obtain Lorentz invariant expression for the collision time
directly from Eq.~(\ref{eq:time}), but we can start from the
covariant equation of motion.
By using the relations $E^*_1 = (P\cdot p_1)/|P|$ and
$t_c^* = (P\cdot x_c)/|P|$,
Eq.~(\ref{eq:motion}) can be rewritten by using 4-vectors:
\begin{align}
 x_1' = x_1 + \frac{P\cdot (x_c - x_1)}{P\cdot p_1}p_1,\nonumber\\
 x_2' = x_2 + \frac{P\cdot (x_c - x_2)}{P\cdot p_2}p_2,\nonumber
\end{align}
where $x_1'$ and $x_2'$ denote the position of the closest approach.
Using the condition $(x_1'-x_2')\cdot(p_1-p_2)=0$,
one can get the collision times as
\begin{align}
 \frac{P\cdot(x_c-x_1)}{P\cdot p_1} 
   &= \frac{(P\cdot x)(p_2\cdot p)-(p\cdot x)(P\cdot p_2)}
           {(p\cdot p_1)(P\cdot p_2)-(p\cdot p_2)(P\cdot p_1)},\nonumber\\
 \frac{P\cdot(x_c-x_2)}{P\cdot p_2} 
  &= \frac{(P\cdot x)(p_1\cdot p)-(p\cdot x)(P\cdot p_1)}
           {(p\cdot p_1)(P\cdot p_2)-(p\cdot p_2)(P\cdot p_1)}.
\end{align}
Two colliding particles are propagated to each collision point
of the closest approach:
\begin{align}
 x_1(t_c) &= x_1(t_1) +\frac{p_2^2(x\cdot p_1) - (p_1\cdot p_2)(x\cdot p_2)}
                            {(p_1p_2)^2 - p_1^2p_2^2} p_1,\\
 x_2(t_c) &= x_2(t_2) -\frac{p_1^2(x\cdot p_2) - (p_1\cdot p_2)(x\cdot p_1)}
                            {(p_1p_2)^2 - p_1^2p_2^2} p_2.
\end{align}
The two collision times are in general different, because of the 
finite spatial separation. Therefore 
one needs a prescription to choose the
time of the ordering of each collision.
In the model, we assume that the collisions are ordered by the
average time $t_\text{order}=(t_{1, \mathrm{col}}+t_{2, \mathrm{col}})/2$
in the computational frame.
Other choice is to use the  time of closest approach for the
two colliding particles in the computational frame as used in UrQMD,
\begin{equation}
  t_{\mathrm{coll}} =  - \frac{(\bm{x}_1-\bm{x}_2)\cdot (\bm{v}_1 - \bm{v}_2)}
         {(\bm{v}_1 - \bm{v}_2)^2}
\end{equation}
The effect of different definition of ordering time was
investigated in Ref.~\citen{Zhang:1996gb}.
The problem of superluminous signals has been studied
in Ref.~\citen{Kortemeyer:1995di}

In this geometrical method for two body collision, collision
occurs in the separated points, which is the same as action-at-distance
interaction. In order to solve this problem, one can use the
``full-ensemble" method or
``subdivision" technique\cite{Welke:1989dr,Molnar:2000jh,Xu:2004mz}
in which cross section is reduced by the factor of the number
of test (over sampling) particle $\sigma/N_\text{test}$ 
in order to recover the local nature of Boltzmann collision term
in the limit of $N_\text{test}\to \infty$.
However, this method is in general computationally expensive
for large $N_\text{test}$. A faster method which is called
``local-ensemble" method has been proposed
in Ref.~\citen{localensemble}.

In order to recover the problem of collision ordering,
the stochastic method~\cite{Danielewicz:1991dh,localensemble,Xu:2004mz}
can be used in which probability is used
to determine the collision instead of geometrical interpretation.
The collision probability for two particle collision  during the time interval
$\Delta t$ in the volume element $V$ is given by
\begin{equation}
 P = v_\text{rel}\frac{\sigma}{N_\text{test}}\frac{\Delta t}{V}
\end{equation}
where $v_\text{rel}$ is the relative velocity of the scattering particles.
In the limit of $N_\text{test}\to\infty, \Delta t \to 0, V \to 0$,
the stochastic algorithm will converge to the exact solutions
of Boltzmann equation, and as a result, recover the Lorentz invariance.
Inclusion of three body collision in the stochastic method
is straightforward.

\subsection{Cross sections}
\label{sec:xsec}
In this section, we summarize modeling of various hadron-hadron ($hh$)
collisions in JAM.
the inelastic $hh$ collisions produce resonances at low energies
   while at high energies
  color strings are formed and they decay into hadrons according to the
  Lund string model with some formation time.
 Formation point and time are determined
  by assuming yo-yo formation point~\cite{bialas}.
 This gives roughly formation time of 1 fm/$c$.

\subsubsection{Resonance productions}
In our approach, it is necessary to input various hadron-hadron ($hh$) cross
sections. We use parameterized cross section for total and elastic collisions.
Total hadronic cross section in JAM is divided in general by
various processes:
\begin{equation}
 \sigma_\text{tot}(s) = \sigma_\text{el}(s) +
 \sigma_\text{ch}(s)+\sigma_\text{ann}(s)
 + \sigma_\text{t-R}(s) + \sigma_\text{s-R}(s)
 + \sigma_\text{t-S}(s) + \sigma_\text{s-S}(s),
\end{equation}
where $\sigma_\text{el}(s), \sigma_\text{ch}(s)$ and $\sigma_\text{ann}(s)$
denote the elastic, charge exchange and annihilation cross sections,
respectively.
$\sigma_\text{t-R}(s)$ and $\sigma_\text{s-R}(s)$
are the $t$ and $s$-channel resonance production cross sections,
such as $NN \to N\Delta$ or $\pi N \to \Delta$,
and
$\sigma_\text{t-S}(s)$ and $\sigma_\text{s-S}(s)$
are the $t$ and $s$-channel string formation cross sections.

Particle production in $hh$ collision is modeled by the
resonance formation for discrete resonance region,
\begin{equation}
 h_1 h_2 \leftrightarrow h_1 h_2^*,~~
 h_1 h_2 \leftrightarrow h_1^* h_2^*,
 ~~~\text{or} ~~~~~ h_1 h_2 \leftrightarrow h_3^*.
\end{equation}
Resonance regions for baryon-baryon ($BB$), baryon-meson ($BM$)
and meson-meson collisions are assumed to be
$\sqrt{s}\leq 4, 3$ and 2 GeV, respectively.
In JAM, cross sections of various resonance formation in $BB$ are
parameterized to reproduce pion multiplicities.
In UrQMD~\cite{urqmd}, BUU~\cite{Teis:1996kx},
matrix elements are fitted to the available data for pion
production cross sections.

The cross section for the inverse process
such as $h_1^* h_2^* \to h_1 h_2$ can be obtained by the detailed balance
formula~\cite{Danielewicz:1991dh,Wolf2,detbal2}
 which takes the finite width of the resonance mass into account.
The differential cross section for the reaction $(3,4)\to(1,2)$
can be expressed by the cross section for $(1,2)\to(3,4)$;
\begin{equation}
  {d\sigma_{34\to 12}\over d\Omega}=
       {(2S_1+1)(2S_2+1)\over (2S_3+1)(2S_4+1)}
          {p^2_{12} \over p_{34}}
    {d\sigma_{12\to 34}\over d\Omega}
     {1\over \int\int p_{34} A(m_3)A(m_4)d(m_3^2)d(m_4^2)}~.
\end{equation}
where  $S_i$ denotes the spin of incident or outgoing particles.
Mass distribution function $A(m_i^2)$ for resonances 
with the mass $m_R$ is given
by the relativistic Breit-Wigner function
\begin{equation}
  A(m^2)={1\over \cal N}{m_R\Gamma(m)\over (m^2-m_R^2)^2 + m_R^2\Gamma(m)^2}.
 \label{eq:rlorentz}
\end{equation}
where $\cal N$ denotes the normalization constant.

The resonance formation cross section for $MB$ and $MM$ collisions
is computed by
using the Breit-Wigner formula~\cite{Brown,rqmd1}
(neglecting the interference between resonances),
\begin{eqnarray}
    \label{eq:bw}
\sigma(MB \to R)
& =  &{\pi(\hbar c)^2 \over p_{\mathrm{cm}}^2}
         \sum_{R}  |C(MB,R)|^2  \nonumber \\
& \times & {(2S_R+1) \over (2S_M+1)(2S_B+1)}
                     {\Gamma_R(MB)\Gamma_R(\mathrm{tot}) \over
                      (\sqrt{s}-m_R)^2+\Gamma_R(\mathrm{tot})^2/4} \ .
\end{eqnarray}
$S_R$, $S_B$ and $S_M$ denote the spin of the resonance,
the decaying baryon and meson respectively.
The sum runs over resonances,
   $R=N(1440)$ - $N(1990)$, $\Delta(1232)$ -$\Delta(1950)$,
   $\Lambda(1405)$ - $\Lambda(2110)$,
   $\Sigma(1385)$ - $\Sigma(2030)$
   and $\Xi(1535)$ - $\Xi(2030)$.
Actual values for these parameters are 
 taken from the Particle Data Group~\cite{pdg1996}
 and adjusted
 within an experimental error bar
to get reasonable fit for $MB$ cross sections.
The momentum dependent decay width is used for the calculation of
the decay width in Eq.~(\ref{eq:bw}),
\begin{equation}
 \Gamma_R(MB)=\Gamma^0_R(MB) {m_R\over m}
              \left({p_{\mathrm{cms}}(m)\over p_{\mathrm{cms}}(m_R)}\right)^{2\ell+1}
   {1.2 \over 1+0.2\left({p_{\mathrm{cms}}(m)\over p_{\mathrm{cms}}(m_R)}\right)^{2\ell+1}}
   \label{eq:width}
\end{equation}
where $\ell$ and $p_{\mathrm{cms}}(m)$ are
 the relative angular momentum
 and 
 the relative momentum in the exit channel in their rest frame.
The Breit-Wigner formula Eq.~(\ref{eq:bw}) is used for
meson-meson collision for resonance production as well.
Meson resonance states are included up to about 1800 MeV.

When there is no available experimental data for cross section,
we use additive quark model~\cite{rqmd1,urqmd},
\begin{equation}
 \sigma_{\mathrm{tot}} = \sigma_{NN}\frac{n_1}{3}\frac{n_2}{3}
     \left(1-0.4\frac{n_{s1}}{n_1}\right)
     \left(1-0.4\frac{n_{s2}}{n_2}\right) \ ,
\end{equation}
where $\sigma_{NN}$ is the total nucleon-nucleon cross section,
$n_i$ is the number of constituent quarks in a hadron, and
$n_{si}$ is the number of strange quarks in a hadron.
Cross sections involving hadrons with many strange quarks
such as $\phi=(s\bar{s})$ or $\Omega=(sss)$ 
are suppressed.
Interestingly, this leads to violation of mass ordering in
differential elliptic flow, \textit{e.g.}, for $\phi$ mesons.~\cite{Hirano:2007ei}
This expression is a good approximation above the resonance region
where cross section becomes flat.
Additive quark cross section yields $\sigma_{K^- p} \approx 21$ mb and
$\sigma_{\Lambda p} \approx 35$ mb, which are consistent with the experimental data. 

\subsubsection{String formation} \label{sec:StringFormation}

At an energy range above $\sqrt{s}>4$-$5$ GeV,
the (isolated) resonance picture breaks down because
width of the resonance becomes wider and 	
the discrete levels get closer.
The hadronic interactions at
the energy range 4-5 $<\sqrt{s}<$ 10-100 GeV where
it is characterized by the small transverse momentum transfer
is called ``soft process", and
string phenomenological models
are known to describe the data for such soft interaction well.
The hadron-hadron collision 
leads to a string like excitation longitudinally.
In actual description of the soft processes,
we employ the prescription adopted in the HIJING model~\cite{hijing}
to treat soft excitation processes.
In HIJING or FRITIOF~\cite{fritiof}, 
excited strings after interaction have the same quark contents
unlike the model based on Gribov-Regge models such as Dual Parton
Models (DPM)~\cite{dpm} or the VENUS model~\cite{venus}.
In DPM or VENUS, strings typically have different quark content
than original hadrons as a result of color exchange.

We shall review string excitation employed by HIJING.
In the center of mass frame of two colliding hadrons,
we introduce light-cone momenta 
$p^{\pm} = E \pm  p_z$.
Assuming that beam hadron 1 moves in the positive $z$-direction
and target hadron 2 moves negative $z$-direction,
the initial momenta of the both hadrons are
\begin{equation}
  p_1 = (p_1^+,p_1^-,0_T), \qquad p_2 = (p_2^+,p_2^-,0_T)\ . 
\end{equation}
After exchanging the light-cone momentum
$(q^+,q^-,\bm{p}_T)$, the momenta will change to
\begin{equation}
       p'_1 = (p_1^+  +p_2^+ - p^+_f,\  p^-_f,\  \bm{p}_T),
 \qquad p'_2 = (p_f^+,\ p_1^- + p^-_2 - p_f^-,\ -\bm{p}_T), 
\end{equation}
where final momenta $p^+_f$ and $p^-_f$ are related to the
longitudinal momentum transfer $q^{\pm}$ as
\begin{equation}
  p^+_f = q^+ + p_2^+, \qquad p^-_f = q^- + p^-_1.
\end{equation}
Using light cone momentum transfer $x^{\pm}$ defined by
\begin{equation}
 x^+ = \frac{p_f^+}{\sqrt{s}}, \qquad x^- = \frac{p_f^-}{\sqrt{s}},
\end{equation}
Final momenta are given by
\begin{equation}
       p'_1 = ((1-x^+)\sqrt{s},\  x^-\sqrt{s},\  \bm{p}_T),
 \qquad p'_2 = (x^+\sqrt{s},\  (1-x^-)\sqrt{s},\ -\bm{p}_T).
\end{equation}
Thus the string masses will be
\begin{equation}
  M_1^2= x^-(1-x^+)s-p^2_T, \qquad M_2^2=x^+(1-x^-)s-p^2_T,
\end{equation}
respectively.
Minimum momentum fractions are
$x_\text{min}^+=p_2^+/P^+$ and $x_\text{min}^-=p_1^-/P^-$.
For the probability for light-cone momentum transfer
in the non-diffractive events,
 we use the same distribution as that in HIJING~\cite{hijing}:
\begin{equation}
  P(x^{\pm})= {(1.0-x^{\pm})^{1.5}\over (x^{\pm2}+c^2/s)^{1/4}}
\end{equation}
for baryons and
\begin{equation}
  P(x^{\pm})= {1\over (x^{\pm2}+c^2/s)^{1/4}((1-x^{\pm})^2+c^2/s)^{1/4}}
\end{equation}
for mesons, where $c=0.1 $GeV is a cutoff.
For single-diffractive events, in order to reproduce
experimentally observed mass distribution $dM^2/M^2$,
we use the distribution
\begin{equation}
   P(x^{\pm})={1 \over (x^{\pm2}+c^2/s)^{1/2}}.
\end{equation}

The same functional form as the HIJING model~\cite{hijing}
  for the soft $\bm{p}_T$ transfer at low 
$p_T<p_0$
is used
\begin{equation}
   f(\bm{p}_T) = \left\{ (p_T^2+c_1^2)(p_T^2+p_0^2)
                     (1+e^{(p_T-p_0)/c_2})  \right\}^{-1} ~,
\end{equation}
where $c_1=0.1$ GeV/$c$, $p_0=1.4$ GeV/$c$ and $c_2=0.4$ GeV/$c$,
to reproduce the high momentum tail of the particles
at energies $E_{\mathrm{lab}}=10$ - $20$ GeV.

\subsubsection{String decay} \label{sec:StringDecay}

The strings are assumed to hadronize via quark-antiquark 
creation using Lund fragmentation model PYTHIA6.1\cite{pythia}.
Hadron formation points from a string fragmentation
are assumed to be given by
  the yo-yo formation point~\cite{bialas} which is defined by the 
  first meeting point of created quarks.
Yo-yo formation time is about 1 fm/$c$ assuming the string
tension $\kappa=1$ GeV/fm.

\begin{figure}[htb]
\begin{center}
\includegraphics[width=3.4in]{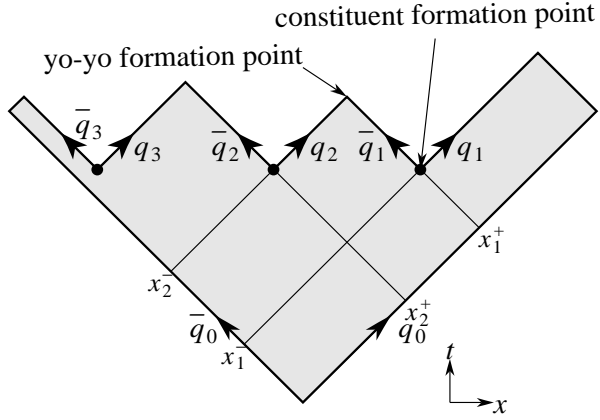}
\end{center}
\caption{Space-time picture of the $q\bar{q}$ string motion and
the definition of formation points.}
\label{fig:1}
\end{figure}

In the Lund string model, space-time coordinates and energy-momentum
coordinates for the quarks are directly related
 via the string tension~\cite{lund,lund2}.
Let us consider one-dimensional massless $q\bar q$ string in the c.m. frame.
If $x_i^{\pm}=t_i\pm x_{zi}$ denote the light-cone coordinates of the 
$i$-th production point, then the light cone momenta
$p_i^{\pm}=E_n\pm p_{zi}$ of the $i$-th rank hadron
which is produced by the energy-momentum fraction $z_i$
from ($i-1$)-th string $  p^+_i = z_i p^+_{i-1}$
 are fixed by
\begin{equation}
p_i^+ = \kappa (x_{i-1}^+ - x_i^+),\quad
p_i^- = \kappa (x_i^- - x_{i-1}^-),
\end{equation}
with initial value for quark $q_0$ moving to the right
\begin{equation}
  x^+_0 = {W\over \kappa},\qquad x^-_0=0,
\end{equation}
where $W$ corresponds to the string initial invariant mass.
The probability distribution for the momentum fraction $z_i$ is 
given by the Lund symmetric fragmentation function
\begin{equation}
f(z) \propto \frac{(1-z)^a}{z}
           \exp\left(-b\frac{(m^2+\bm{p}_\perp^2)}{z}\right) \ ,
\end{equation}
where $a$ and $b$ are parameters which have to be fitted to 
experimental data, and $m$ and $\bm{p}_\perp$ denote the mass
and transverse momentum of the produced hadron, respectively.
This form can be obtained by the condition of the left-right symmetry
for the decay of string.

Using the relation $p_i^+p_i^-=m_{i\perp}^2$ with
$m_{i\perp}$ being transverse mass of the $i$-th hadron,
we have the recursion formulae~\cite{lund}
\begin{equation}
  x_i^+ = (1-z_i)x^+_{i-1}, \quad
  x_i^- = x^-_{i-1} + \left({m_{i\perp}\over \kappa}\right)^2
                   {1-z_i\over z_i}{1\over x^+_i}.
\end{equation}
Yo-yo formation points where the two $q\bar{q}$ meet
for the first time are obtained as
\begin{equation}
    x_i^{\mathrm{yo-yo}}=(x_{i-1}^+,x_i^-), \label{eq:yoyo}
\end{equation}
and constituent formation points where $q\bar{q}$ is created
\begin{equation}
    x_i^{\mathrm{const}}=(x_i^+,x_i^-). \label{eq:const}
\end{equation}
In RQMD~\cite{rqmd2}, the formation points of hadrons are calculated
as the average of the two $q\bar q$ production point as
\begin{equation}
    x_i^{\mathrm{RQMD}}
  =\left({x_i^+ + x_{i-1}^+\over 2},{x_i^-+x_{i-1}^-\over 2}\right).
  \label{eq:rqmd}
\end{equation}
Clearly, one can see that
\begin{equation}
    x_i^{\mathrm{const}} < x_i^{\mathrm{RQMD}} < x_i^{\mathrm{yo-yo}} ~.
\end{equation}
It is assumed in UrQMD and JAM that the yo-yo formation point is assigned to the space-time
point for produced hadrons.
The produced hadrons which have the original constituent quarks are
allowed to scatter with reduced cross section according to the
number of constituent quarks. For example, if baryon has one original
constituent quark, cross section during the formation time
would be reduced by the factor one-third.

\section{Non-Abelian plasmas}
\label{sec:nonabelian}

Understanding the dynamics of the non-equilibrium system
and the process toward the thermalization is a very important outstanding
question in high energy heavy ion collisions.
In Ref.~\citen{BMSS}, ``Bottom-up thermalization'' was proposed
based on the Boltzmann equation with inelastic processess.
Later, full numerical simulations of the Boltzmann equation~\cite{Xu:2004mz}
show that inelastic gluon scatterings are very important processes
for thermalization of gluonic systems.
It was, however, pointed out that the soft classical color fields
may play an important role in the dynamics of thermalization.
Specifically, Weibel-like QCD plasma instabilities may develop due to
anisotropic distribution of hard plasma particles
(partons)~\cite{Mrowczynski,ALM,RomStri}.
First numerical simulation was done within the hard-loop approximation
by solving linearized Vlasov equation in one-dimension~\cite{RRS05}.
In this analysis, exponential growth of gauge fields due to Abelianization
of the field was found.
In full 3-dimensional simulations~\cite{AMY3d,Rebhan:2005re}, 
it was found that instabilities
grow linearly once the field strength becomes large and
non-Abelian self-interactions among gluon fields become
nonperturbative in the case of moderate anisotropic momentum distribution.
In this case,
non-Abelian interactions develop a cascade of energy from soft modes
to hards modes~\cite{AM3}.
However, in the case of extreme anisotropy~\cite{Dumitru:2006pz,Bodeker:2007fw},
energy coming from plasma particles due to Weibel-like instability
does not go to the soft mode, instead go back to the hard scale
very rapidly which is called ``ultraviolet avalanche"  in non-Abelian
plasmas.

Instabilities of the gluon fields have been also investigated
by employing pure classical Yang-Mills
equation~\cite{Romatschke:2005pm,Fukushima:2006ax,Iwazaki:2007es,Berges:2007re}.
It was pointed out~\cite{Iwazaki:2007es} that
instabilities in the classical Yang-Mills field are the Nielsen-Olsen
type instability. Numerical analysis was performed in
Ref.~\citen{Berges:2011sb}.
In Nielsen-Olsen instability, unstable mode
exists at zero momentum, while zero momentum mode is stable in
Weibel instability. This is one of the crucial differences between Weibel
and Nielsen-Olsen instabilities.

In this section, we review how to solve Wong-Yang-Mills equation based on the technique
developed in Ref.~\citen{Dumitru:2006pz} below.

\subsection{Simulation of non-Abelian plasma}
\label{sec:vlasov}

We present numerical method of
the classical Vlasov transport equation for gluons with
non-Abelian color charge $q^a$ 
\begin{equation}
p^{\mu}[\partial_\mu + gq^aF^a_{\mu\nu}\partial^\nu_p +
gf_{abc}A^b_\mu q^c\partial_{q^a}]f(x,p,q)=0 \label{eq:vlasov} 
\end{equation}
where $f(\bm{x},\bm{p},q)$ denotes the single-particle distribution
function and $g$ is the gauge coupling.
The Vlasov equation is coupled self-consistently to the
Yang-Mills equation
\begin{equation}
 D_\mu F^{\mu\nu} = J^\nu= g\int\frac{d^3p}{(2\pi)^3} dq qv^\nu
    f(t,\bm{x},\bm{p},q),
\end{equation}
with $v^\nu=(1,\bm{p}/p)$, $D_\mu = \partial_\mu +ig A_{\mu}$
$F_{\mu\nu}=\partial_\mu A_\nu - \partial_\nu A_\mu +ig[A_\mu,A_\nu]$, and
$J^\nu$ denotes the current generated by plasma particles.
This set of equations reproduces the ``hard thermal loop'' effective
theory~\cite{ClTransp} near equilibrium.

The Vlasov equation can be solved by replacing the one-particle
distribution function with 
many ``test particles":
\begin{equation}
 f(\bm{x},\bm{p},q) = \frac{(2\pi)^3 }{N_{\mathrm{test}}}\sum_i 
     \delta(\bm{x}-\bm{x}_i(t)) 
         \delta(\bm{p}-\bm{p}_i(t))
        \delta(q-q_i(t)),
\end{equation}
where $N_\mathrm{test}$ is the number of test particles,
and $\bm{x}_i(t)$, $\bm{p}_i(t)$ and $q_i(t)$ are
the coordinate, momentum and the color charge of the test particle.
From  this assumption, one gets the Wong equation~\cite{Wong}
for the $i$-th ``test particle"
\begin{equation}
\frac{d\bm{x}_i}{dt} = \bm{v}_i ,\quad
\frac{d\bm{p}_i}{dt} = g\, q_i^a \,
\left( \bm{E}^a + \bm{v}_i \times \bm{B}^a \right),
\quad \frac{dq_i}{dt} = -ig\, v^{\mu}_i \, [ A_\mu, q_i]
\label{eq:wong}
\end{equation}

The time evolution of the Yang-Mills field can be
followed by the Hamiltonian method~\cite{Ambjorn} in the temporal gauge.
The Hamilton equation of motion for gauge field $A_i=A_i^at^a$
and the color electric field $E_i=E_i^at^a$
under the temporal gauge $A^0=0$ is
\begin{equation}
  \frac{dA_i}{dt} =  E^i, \qquad
  \frac{dE^i}{dt} = \sum_j D_jF_{ji} -J^i, \quad (i,j=x,y,z)
\end{equation}

One may descritize the equation on the lattice size $a$
by introducing the link variable $U_i(x)=\exp(iagA_i(x))$
\begin{eqnarray}
\dot{U}_i(x) &=&  iagE^i(x)U_i(x), \\
\dot{E}^i_a(x) &=& \frac{-i}{2a^3g}
\sum_j\Tr
\left\{ 
    t^a[ U_{ji}(x)-U_j^\dagger(x-j)U_{ji}(x-j)U_j(x-j)   ] 
                        \right\}   -J^i_a
\end{eqnarray}
where, the plaquette is defined as
$U_{\Box}= U_{ij}(x)=U_i(x)U_j(x+i)U^\dagger_i(x+j)U^\dagger_j(x)$,
and $t^a=\sigma^a$ is the Pauli matrix for SU(2) case.
This equation is covariant under the lattice gauge transformation
\begin{equation}
  U_i(x) \to V(x)U_i(x)V^\dagger(x+i), \qquad
  E^i(x) \to V(x)E^i(x)V^\dagger(x), \qquad
\end{equation}

Typically one uses the time step size of $\Delta t\approx 0.05a$
to ensure energy conservation
\begin{equation}
H = \sum_{i,x} {a^3\over2} E^2_i(x)
   +\sum_{\Box}
    {1\over2g^2a}(N_c - \RE\Tr U_{\Box})
    + \frac{1}{N_\text{test}}\sum_j |\bm{p}_j|
\end{equation}
and Gauss law
\begin{equation}
\sum_i \left(E_i(x) - U_i^\dagger(x-i)E_i(x-i)U_i(x-i) \right)= \rho(x)
\end{equation}

Numerical method has been developed to solve the equations~(\ref{eq:wong})
in Ref.~\citen{HuMullerMoore} which is the non-Abelian extension
of the nearest-grid-point (NGP) method.
In the NGP method, the charge density is obtained by counting
the number of particles within a cell.
A current $J(x)=Q\delta(t-t_\text{cross})/N_\text{test}$ is generated
only when a particle with the color charge $Q$ crosses a cell boundary
from $x$ to $x+i$ at the time $t_\text{cross}$.
In order to satisfy the requirement of the lattice covariant continuity
equation 
\begin{equation}
  \dot{\rho}(x) = \sum_i U_i^\dagger(x-i)J_i(x-i)U_i(x-i) - J_i(x)\ ,
\end{equation}
the color charge must be parallel transported
\begin{equation}
 Q(x+i) = U_i^\dagger(x)Q(x)U_i(x)\ .
\end{equation}
At each time step, an effect of magnetic field, 
which causes a rotation of momentum and
does not change the energy of the particle, on the particle
is taken into account. 
But the magnitude of the momentum can change when it crosses a cell
boundary. The final momentum $|\bm{p}_\mathrm{fin}|$ 
after crossing the cell can be
fixed by the energy conservation
\begin{equation}
 |\bm{p}_\mathrm{ini}| + \frac{N_\mathrm{test}}{2}\bm{E}_\mathrm{ini}^2
    =
    |\bm{p}_\mathrm{fin}| + \frac{N_\mathrm{test}}{2}
       (\bm{E}_\mathrm{ini} - J )^2~.
\end{equation}
Substituting $J={Q}/{N_\mathrm{test}}$ and  when
a particle crosses the cell boundary in $x$-direction, 
one obtains for new momentum in the $x$-direction
\begin{equation}
 p_{x} = \sqrt{|\bm{p}_\mathrm{ini}|
        - E_{x,\mathrm{ini}}Q + Q^2/(2N_\mathrm{test}) - p_\perp^2}~,
\end{equation}
where $|\bm{p}_\mathrm{fin}| = \sqrt{p_x^2 + p_\perp^2}$ is used.
As we explained above, because of the current discontinuities introduced 
in NGP method, large amounts of noise are generated. In order to
eliminate this noise, we need a large number of test particle.
NGP method was applied in~\cite{Dumitru:2005gp,Dumitru:2005hj,Nara:2005fr} 
for one-dimensional case in which gauge field is assumed to
depend only on one-direction: $\bm{A}(x,y,z)=\bm{A}(z)$.
We call this approximation as 1d-3v simulation because particle velocity
has three direction.
In 1d-3v case, we can take a very large number of test particle
to obtain sufficiently smooth current. 
However, for full 3d-3v simulation, computational cost may be
very expensive due to the larger number of test particle
required, and we need improved algorithms.

\subsubsection{Particle-in-Cell (PIC) simulations}

In Abelian plasmas, particle-in-cell (PIC) method is widely 
used in which smoothed currents are employed~\cite{Birdsall,Hockney,kembo1}
in order to suppress numerical noise.
The $m$-th order shape function (b-spline) $S_m$ can be obtained
by the convolution
of $S_{m-1}$ with the nearest-grid-point weighting function $S_0$
\begin{equation}
  S_m(x) = \int^{\infty}_{-\infty} S_0(x') S_{m-1}(x-x')dx'
  \label{eq:convolution}
\end{equation}
where zero-th-order function is the flat-top function
\begin{equation}
 S_0(x) = \left\{
  \begin{array}{ll}
    1              & \mathrm{for}\, |x| \leq \dfrac{1}{2},\\
      0            & \mathrm{otherwise}
  \end{array}
    \right.
\end{equation}
Commonly used is the first-order function $S_1$ which corresponds
to linear interpolation
\begin{equation}
 S_1(x) = \left\{
  \begin{array}{ll}
     1-|x|         & \mathrm{for}\, |x| \leq 1,\\
      0            & \mathrm{otherwise}
  \end{array}
    \right.
\end{equation}
The charge density at each lattice site is given by the superposition
of a particle,
\begin{equation}
 \rho(i,j,k) = \sum_\text{particle} \frac{q}{\Delta x\Delta y\Delta z}
           S_m\left(\frac{x-i}{\Delta x}\right)
           S_m\left(\frac{y-i}{\Delta y}\right)
           S_m\left(\frac{z-i}{\Delta z}\right),
\end{equation}
where $(x,y,z)$ is the coordinate of particle, and
$\Delta x$, $\Delta y$ and $\Delta z$ are the lattice spacing in the
$x$, $y$ and $z$ directions, respectively.
However, discrete continuity equation would not necessarily be satisfied 
on the lattice. One may need to solve Poisson equation to
recover Gauss law.
More efficient numerical methods that
satisfy exactly the lattice continuity equation have been
developed~\cite{Eastwood,Bueman,Umeda}.
In PIC simulation, the total amount of charge is distributed over its
surface, and each charge contributes to the charge density in several
cells (8 cells for first order in 3-dimension).
In NGP method, current is generated only when a particle crosses
the cell boundary, but in PIC, current will be continuously generated
in a given time interval
which  is the amount of charge crossing a cell boundary.

Let us try to construct a current which satisfies the lattice
continuity equation.
Difference of charge density within a time step $\Delta t$
at a lattice site $(i,j)$ in 2-dimensional case is
\begin{align}
& \rho^{t+\Delta t}(i,j) -\rho^{t}(i,j)\nonumber \\
 &= \sum_\text{particle}
      \frac{q}{\Delta x\Delta y}
      \left\{
           S^m_i(x(t+\Delta t))S^m_i(y(t+\Delta t))
          -S^m_i(x(t))S^m_i(y(t))
	  \right\}
	  \nonumber\\
 &= \sum_\text{particle} \int_{t}^{t+\Delta t} \frac{d}{dt}
            \left (qS_i^m(x(t)) S_i^m(y(t)) \right) dt\ ,
\end{align}
where $S_i^m(x)=\frac{1}{\Delta x}S_m\left(\frac{x-i}{\Delta x}\right)$.
By using the identity for the derivative of $S_m$
\begin{equation}
 \frac{dS_m(x/\Delta x)}{dx} = \frac{1}{\Delta x}[
         S_{m-1}(x+\Delta x/2) -S_{m-1}(x-\Delta x/2)]
\end{equation}
which follows from the Eq.~(\ref{eq:convolution}),
one finds that the current  defined as
\begin{equation}
  J_\alpha(i,j,k)= \frac{1}{\Delta x\Delta y\Delta z}\sum_\text{particle}
           q v_\alpha S_{m-1}\left(\frac{x-i-\Delta x/2}{\Delta x}\right)
	              W_m\left(\frac{y-j}{\Delta y}\right)
		      W_m\left(\frac{z-k}{\Delta z}\right)
\end{equation}
satisfies the lattice continuity equation,
where
\begin{equation}
 W_m(x) = \int^{t+\Delta t}_t S_m(x) dt\ .
\end{equation}

We consider a particle moving from $(x_1, y_1)$ to $(x_2, y_2)$
during a time step $\Delta t$, \textit{i.e.},
$x_2= x_1 + v_x\Delta t, \quad y_2= y_1 + v_y\Delta t$.
We also restrict ourselves to the case in which
$(x_2,y_2)$  belongs to the same lattice site $(i,j)$.
In this case,
The explicit form for the first shape-factor corresponding
linear smearing case is given by
\begin{eqnarray}
\rho(i,j) &=& \frac{1}{\Delta x\Delta y}(1-x)(1-y),
     \quad \rho(i,j+1) = \frac{1}{\Delta x\Delta y}x(1-y)~, \nonumber\\
\rho(i,j) &=& \frac{1}{\Delta x\Delta y}(1-x)y, 
     \qquad \rho_y(i+1,j+1) = \frac{1}{\Delta x\Delta y}xy, \\
J_x(i,j) &=& \frac{1}{\Delta x\Delta y}F_x(1-W_y), 
      \qquad J_x(i,j+1) = \frac{1}{\Delta x\Delta y}F_xW_y~, \nonumber\\
J_y(i,j) &=& \frac{1}{\Delta x\Delta y}F_y(1-W_x), 
       \qquad J_y(i+1,j) = \frac{1}{\Delta x\Delta y}F_yW_x~, \label{eq:smCurrent}
\end{eqnarray}
where $x=(x_1-i)/\Delta x$, $y=(y_1-j)/\Delta y$,
and $(F_x,F_y) \equiv \bm{F}$ represents the charge flux
\begin{equation}
F_x = q\frac{x_2-x_1}{\Delta t}~, \qquad
F_y = q\frac{y_2-y_1}{\Delta t}~.
\end{equation}
$W_{x/y}$ is defined at the midpoint between
the starting point $(x_1,y_1)$ and the end point $(x_2,y_2)$
\begin{equation}
  W_x = \frac{x_1+x_2}{2}-i~,\qquad
  W_y = \frac{y_1+y_2}{2}-j~.
\end{equation}
We use the `Zigzag scheme' developed in Ref.~\citen{Umeda}
in case particle crosses the cell boundary within a time step.

Finally, we note that the electromagnetic forces should be smeared in
a similar way when a particle momentum is
updated~\cite{Eastwood}. 
Consider the time derivative of the total energy:
\begin{equation}
 \frac{dE_\text{tot}}{dt} = -\sum_\text{lattice} \bm{E}\cdot \bm{J}
             + \sum_\text{particles} q_i \, \bm{E}(\bm{x}_i(t))\cdot \bm{v}_i.
\end{equation}
It is clear that the interpolation function for $\bm{E}$ should be the
same as that for $\bm{J}$ in order to achieve good energy conservation
in the simulation.  The electric field $\bm{E}(\bm{x})$ at the particle
position $\bm{x}$ is then obtained from
\begin{equation}
 E_\alpha(\bm{x}) = \sum_{\text{lattice}}
         S^0_\alpha S^1_\beta S^1_\gamma E_\alpha(i,j,k)~,
\end{equation}
while the magnetic field is given by
\begin{equation}
 B_\alpha(\bm{x}) = \sum_{\text{lattice}}
         S^1_\alpha S^0_\beta S^0_\gamma B_\alpha(i,j,k)~.
\end{equation}
This is motivated by the relations $\bm{E} =-\nabla A^0-\dot{\bm{A}}$ and
$ \bm{B} = \nabla \times \bm{A}$.

Next we consider the discretization of the equation Eq.~(\ref{eq:wong})
for momentum update.
The difference form of the equation is
\begin{equation}
  \frac{\bm{p}(t+\Delta t/2)-\bm{p}(t-\Delta t/2)}{\Delta t} = 
  gq^a\left(\bm{E}^a(t)
    + \frac{\bm{p}(t+\Delta t/2)+\bm{p}(t-\Delta t/2)}{2}
     \times \frac{\bm{B}^a(t)}{e(t)}  \right)~. \label{eq:mom}
\end{equation}
Equation (\ref{eq:mom}) can be solved
by the Buneman-Boris method~\cite{Hockney,Birdsall}
as follows:
\begin{eqnarray}
  \bm{p}(t) &=& \bm{p}(t-\Delta t/2) + \frac{\Delta t}{2}\bm{E}(t),\\
  e(t) &=& |\bm{p}(t)|,\\
  \bm{p}'(t) &=& \bm{p}(t) + \frac{\Delta t}{2}
                 \bm{p}(t)\times\frac{\bm{B}(t)}{e(t)},\\
  \bm{p}_2(t) &=& \bm{p}(t) + \frac{2}{1+(\bm{B}/e(t)\Delta t/2)^2}
          \frac{\Delta t}{2}\bm{p}'(t)
                 \times\frac{\bm{B}(t)}{e(t)},\\
  \bm{p}(t+\Delta t/2) &=& \bm{p}_2(t) + \frac{\Delta t}{2}\bm{E}(t),
\end{eqnarray}
where $\bm{E}\equiv gq^a\bm{E}^a$ and $\bm{B}\equiv gq^a\bm{B}^a$.
This scheme is time reversible and the overall momentum
integration is accurate to second-order in the time step.

\subsection{PIC simulations in non-Abelian gauge theories (CPIC)}
\label{nonAbelianPIC}
An extension of the charge conserved method to
the smearing method in the non-Abelian case
has been proposed in Ref.~\citen{Dumitru:2006pz} in which
the current is defined as
\begin{eqnarray}
 J_x(i,j) &=& Q\frac{x_2-x_1}{\Delta t}(1-W_y),
         \qquad J_x(i,j+1) = Q_{y}\frac{x_2-x_1}{\Delta t}W_y~, \\
 J_y(i,j) &=& Q\frac{y_2-y_1}{\Delta t}(1-W_x), \qquad 
  J_y(i+1,j) = Q_{x}\frac{y_2-y_1}{\Delta t}W_x~, 
\end{eqnarray}
where we define the parallel transport of the charge in two dimension as
\begin{equation}
 Q_x \equiv U_x^\dagger(i,j)QU_x(i,j),\qquad
 Q_y  \equiv  U_y^\dagger(i,j)QU_y(i,j)~.
\end{equation}
One can easily check that
this satisfies the lattice covariant continuity equation,
\begin{equation}
 \dot{\rho}(i,j) = \sum_x U^\dagger_x(i-x)J_x(i-x)U_x(i-x) - J_x(i,j),
\end{equation}
for sites $(i,j), (i+1,j), (i,j+1)$:
\begin{eqnarray}
\dot{\rho}(i,j) &=& J_x(i,j) + J_y(i,j), \label{eq:ch1}\\
\dot{\rho}(i+1,j)&=&
     U^\dagger_x(i,j)J_x(i,j)U_x(i,j) - J_y(i+1,j),\label{eq:ch2}\\
\dot{\rho}(i,j+1) &=&
    U^\dagger_y(i,j)J_y(i,j)U_y(i,j) - J_x(i,j+1), \label{eq:ch3}
\end{eqnarray}
Equations (\ref{eq:ch1}), (\ref{eq:ch2}) and (\ref{eq:ch3}) are
consistent with the following definitions of the charge densities:
\begin{eqnarray}
 \rho(i,j) &=& Q(1-x)(1-y) ~,\\
 \rho(i,j+1) &=& Q_y(1-x)y ~,\\
 \rho(i+1,j) &=& Q_xx(1-y)~.
\end{eqnarray}
However, since a particle's color charge depends on its path, so
does $\rho(i+1,j+1)$ and we are not able to calculate it from the
charge distribution itself. Rather, we directly employ covariant
current conservation to determine the increment of
charge at site $(i+1,j+1)$ within the time-step.
In this way, we can satisfy Gauss's law in the non-Abelian case.

Finally, we have to check that $\Tr(Q^2)$ is conserved by this smearing
method. This is true when the lattice spacing $a$ is small, as the
total charge of a particle is given by
\begin{equation}
 Q_0 = Q(1-x)(1-y) + Q_xx(1-y) + Q_y(1-x)y + [a_pQ_{xy} + (1-a_p)Q_{yx}]xy~,
\end{equation}
where the $a_p$ depend on the path of a particle and
$Q_{xy}=U^\dagger_x(i,j+1)Q_yU_x(i,j+1)$,
$Q_{yx}=U^\dagger_y(i+1,j)Q_xU_y(i+1,j)$.
If we require that $\Tr(Q_0^2)$ be constant, then the
cross terms, for example $\Tr(QQ_x)$, have to vanish.
This is true when $a$ is small, because $\Tr(Q[A,Q])=0$:
\begin{equation}
 \Tr(QQ_x) = \Tr(Q(Q + iga[A_x,Q] + {\cal O}(a^2))) = \Tr(Q^2) +{\cal O}(a^2).
\end{equation}
Therefore, the magnitude of the color charges is conserved for small lattice
size $a$ in our method.

\section{Summary}
\label{s:conclusion}

We have reviewed recent progress on the development of dynamical
models in heavy ion collisions.
Our emphasis was put on the technical aspects of the models.
Initial particle production processes are obtained by the
CGC framework based on the $k_T$ factrization formula.
Vlasov simulation of non-Abelian plasma can be followed by
the non-Abelian extension of particle-in-cell method.
In the locally thermailzed stage, relativistic hydrodynamics
can be used to describe space-time evolution of matter.
Energy loss of energetic partons in the expanding
fluid elements is necessary for the consistent description
of the jet quenching as well as two-particle correlations.
Finally, effects of final state interactions during the hadronic
gas state are also important for the realistic simulation of
heavy ion collisions.

Recently there have been made significant progresses for theoretical approaches.
Viscous hydrodynamic simulations have been performed by many groups~\cite{visc}
to extract transport coefficients such as a ratio of shear viscosity 
to entropy density.

Numerical solutions of the classical Yang-Mills
equations are first employed for the initial condition of 
event-by-event hydrodynamical simulations~\cite{Schenke:2012wb}.
JIMWLK renormalization group evolution equation was used as initial conditions
for the classical Yang-Mills equation~\cite{Lappi:2011ju}
in order to incorporate the rapidity evolution of the probability distribution
for Wilson lines. This approach may give an important initial condition
for hydrodynamic simulations.
SU(2) plasma simulations of Boltzmann-Vlasov equation including
both collision term for hard particles and soft interaction by
classical Yang-Mills fields are performed
in Refs.~\citen{Dumitru:2007rp,Schenke:2008gg,Schenke:2008pm,Schenke:2009jy}.
It was demonstrated that results are independent of the choice of
separation scale between hard and soft modes.
Inclusion of inelastic process is important for the jet energy loss
in the early stages of heavy ion collision before thermalization.

One of the outstanding problems in high energy collisions at RHIC and LHC
is the missing understanding of non-equilibrium dynamics in the
early stages.
Recent progress on the investigations
how non-Abelian plasmas or Glasma approach equilibrium state
in heavy ion collisions can be found, \textit{e.g.}, in 
Refs.~\citen{Gelis:2007pw,Dusling:2010rm,Kurkela:2011ti,Hatta:2011ky}.
Non-equilibrium simulations for these approaches will provide
insight into the mechanism of thermalization in heavy ion collisions.

\section*{Acknowledgments}
The authors acknowledges the fruitful collaboration
with
A. Dumitru,
H.-J. Drescher,
P.~Huovinen,
and M. Strickland.
The work  was partly supported by
Grant-in-Aid for Scientific Research
Nos.~22740151 and 22340052.
The work of Y.N.\ was supported by
Grant-in-Aid for Scientific Research
No.~20540276.

%

\end{document}